\documentclass[twocolumn]{aastex701}
\usepackage{multirow}
\usepackage{xcolor}
\usepackage{amsmath}

\newcommand{\um}{\ensuremath{\mu\mathrm{m}}}
\newcommand{\uJy}{\ensuremath{\mu\mathrm{Jy}}}
\newcommand{\fagn}{\ensuremath{f_\mathrm{AGN}}}
\newcommand{\lbol}{\ensuremath{L_\mathrm{bol}}}

\submitjournal{ApJ}

\shorttitle{MEOW}
\shortauthors{Leung et al.}

\begin{document}

\title{The MIRI Early Obscured-AGN Wide Survey (MEOW):\\A Population of Hidden AGN at $z \gtrsim 5$ Revealed by JWST/MIRI Imaging}

\correspondingauthor{Gene C. K. Leung}
\email{gckleung@mit.edu}

\author[0000-0002-9393-6507]{Gene C. K. Leung}
\affiliation{MIT Kavli Institute for Astrophysics and Space Research, 77 Massachusetts Avenue, Cambridge, MA 02139, USA}
\email{gckleung@mit.edu}

\author[0000-0003-2895-6218]{Anna-Christina Eilers}
\affiliation{MIT Kavli Institute for Astrophysics and Space Research, 77 Massachusetts Avenue, Cambridge, MA 02139, USA}
\affiliation{Department of Physics, Massachusetts Institute of Technology, 77 Massachusetts Avenue, Cambridge, MA 02139, USA}
\email{emailaddress}

\author[0000-0003-4564-2771]{Ryan Endsley}
\affiliation{Department of Astronomy, The University of Texas at Austin, Austin, TX 78712, USA}
\email{emailaddress}

\author[0000-0001-8519-1130]{Steven L. Finkelstein}
\affiliation{Department of Astronomy, The University of Texas at Austin, Austin, TX 78712, USA}
\affiliation{Cosmic Frontier Center, The University of Texas at Austin, Austin, TX 78712, USA}
\email{stevenf@astro.as.utexas.edu}

\author[0000-0002-9921-9218]{Micaela B. Bagley}
\affiliation{Department of Astronomy, The University of Texas at Austin, Austin, TX 78712, USA}
\email{emailaddress}

\author[0000-0001-6813-875X]{Guillermo Barro}
\affiliation{University of the Pacific, Stockton, CA 90340 USA}
\email{gbarro@pacific.edu}

\author[0000-0002-6610-2048]{Anton M. Koekemoer}
\affiliation{Space Telescope Science Institute, 3700 San Martin Drive, Baltimore, MD 21218, USA} 
\email{emailaddress}

\author[0000-0003-4528-5639]{Pablo G. P\'erez-Gonz\'alez}
\affiliation{Centro de Astrobiolog\'{\i}a (CAB), CSIC-INTA, Ctra. de Ajalvir km 4, Torrej\'on de Ardoz, E-28850, Madrid, Spain}
\email{pgperez@cab.inta.csic.es}

\author[0000-0003-3382-5941]{Nor Pirzkal}
\affiliation{ESA/AURA Space Telescope Science Institute, 3700 San Martin Drive, Baltimore, MD, 212129}
\email{emailaddress}

\author[0000-0001-8534-7502]{Bren E. Backhaus}
\affil{Department of Physics and Astronomy, University of Kansas, Lawrence, KS 66045, USA}
\email{bren.backhaus@ku.edu}

\author[0000-0001-8174-6389]{Teodora-Elena Bulichi}
\affiliation{MIT Kavli Institute for Astrophysics and Space Research, 77 Massachusetts Avenue, Cambridge, MA 02139, USA}
\affiliation{Department of Physics, Massachusetts Institute of Technology, 77 Massachusetts Avenue, Cambridge, MA 02139, USA}
\email{emailaddress}

\author[0000-0002-6184-9097]{Jaclyn B. Champagne}
\email{jchampagne@stsci.edu}
\affiliation{Space Telescope Science Institute, 3700 San Martin Drive, Baltimore, MD 21218, USA}
\email{emailaddress}

\author[0000-0003-4922-0613]{Katherine Chworowsky}
\affiliation{Department of Astronomy, The University of Texas at Austin, Austin, TX 78712, USA}
\affiliation{Cosmic Frontier Center, The University of Texas at Austin, Austin, TX 78712, USA}
\email{k.chworowsky@utexas.edu}

\author[0000-0001-7151-009X]{Nikko J. Cleri}
\affiliation{Department of Astronomy and Astrophysics, The Pennsylvania State University, University Park, PA 16802, USA}
\affiliation{Institute for Computational and Data Sciences, The Pennsylvania State University, University Park, PA 16802, USA}
\affiliation{Institute for Gravitation and the Cosmos, The Pennsylvania State University, University Park, PA 16802, USA}
\email{cleri@psu.edu}

\author[0000-0001-5414-5131]{Mark Dickinson}\affiliation{NSF's National Optical-Infrared Astronomy Research Laboratory, 950 N. Cherry Ave., Tucson, AZ 85719, USA}
\email{fakeemail4@google.com}

\author[0000-0003-3310-0131]{Xiaohui Fan}
\affiliation{Steward Observatory, University of Arizona, 933 North Cherry Avenue, Tucson, AZ 85721, USA}
\email{xfan@arizona.edu}

\author[0000-0001-7201-5066]{Seiji Fujimoto}
\affiliation{David A. Dunlap Department of Astronomy and Astrophysics, University of Toronto, 50 St. George Street, Toronto, Ontario, M5S 3H4, Canada}
\affiliation{Dunlap Institute for Astronomy and Astrophysics, 50 St. George Street, Toronto, Ontario, M5S 3H4, Canada}
\email{emailaddress}

\author[0000-0001-9440-8872]{Norman A. Grogin}
\affiliation{Space Telescope Science Institute, 3700 San Martin Drive, Baltimore, MD 21218, USA}
\email{nagrogin@stsci.edu}

\author[0000-0002-5537-8110]{Allison Kirkpatrick}
\affiliation{Department of Physics and Astronomy, University of Kansas, Lawrence, KS 66045, USA}
\email{emailaddress}

\author[0000-0002-8360-3880]{Dale D. Kocevski}
\affiliation{Department of Physics and Astronomy, Colby College, Waterville, ME 04901, USA}
\email{emailaddress}

\author[0000-0002-5588-9156]{Vasily Kokorev}
\affiliation{Department of Astronomy, The University of Texas at Austin, Austin, TX 78712, USA}
\affiliation{Cosmic Frontier Center, The University of Texas at Austin, Austin, TX 78712, USA}
\email{emailaddress}

\author[0000-0003-2366-8858]{Rebecca L. Larson}
\altaffiliation{Giacconi Postdoctoral Fellow}
\affiliation{Space Telescope Science Institute, 3700 San Martin Drive, Baltimore, MD 21218, USA}
\email{rlarson@stsci.edu}  

\author[0000-0003-1581-7825]{Ray A. Lucas}
\affiliation{Space Telescope Science Institute, 3700 San Martin Drive, Baltimore, MD 21218, USA}
\email{lucas@stsci.edu}

\author[0000-0001-9879-7780]{Fabio Pacucci}
\affiliation{Center for Astrophysics $\vert$ Harvard \& Smithsonian, Cambridge, MA 02138, USA}
\affiliation{Black Hole Initiative, Harvard University, Cambridge, MA 02138, USA}
\email{emailaddress}

\author[0000-0001-7503-8482]{Casey Papovich}
\affiliation{Department of Physics and Astronomy, Texas A\&M University, College
Station, TX, 77843-4242 USA}
\affiliation{George P.\ and Cynthia Woods Mitchell Institute for
 Fundamental Physics and Astronomy, Texas A\&M University, College
 Station, TX, 77843-4242 USA}
\email{papovich@tamu.edu}

\author[0000-0003-1282-7454]{Anthony J. Taylor}
\email{anthony.taylor@austin.utexas.edu}
\affiliation{Department of Astronomy, The University of Texas at Austin, Austin, TX 78712, USA}
\affiliation{Cosmic Frontier Center, The University of Texas at Austin, Austin, TX 78712, USA}

\author[0000-0001-6917-4656]{Natalia C. Villanueva}
\affiliation{Department of Astronomy, The University of Texas at Austin, Austin, TX 78712, USA}
\email{nataliavillanueva@utexas.edu}

\author[0000-0001-8835-7722]{Guang Yang}
\affiliation{Nanjing Institute of Astronomical Optics \& Technology, Chinese Academy of Sciences, Nanjing 210042, China}
\email{gyang@niaot.ac.cn}

\begin{abstract}

We present the MIRI Early Obscured-AGN Wide Survey (MEOW), a JWST/MIRI imaging survey designed to detect dust-obscured active galactic nuclei (AGN) across cosmic time, with a particular focus on the high-redshift universe at $z \gtrsim 5$. MEOW observes the GOODS-N and GOODS-S fields with 43 pointings covering 95 arcmin$^2$ with the F1000W and F2100W filters, reaching depths of 0.5 and 3.6 $\mu$Jy ($5\sigma$), respectively. Using spectral energy distribution (SED) modeling combining MEOW photometry with archival HST, JWST/NIRCam, and SCUBA-2 data, we identify a sample of 16 MIRI-selected AGN at $z = 4.5$--$7.2$ (12 spectroscopically confirmed), spanning bolometric luminosities of $L_{\rm bol} = 10^{44.6}$--$10^{46.4}$~erg~s$^{-1}$. Twelve of the 16 AGN are newly identified in this work, including at least five narrow-line AGN representing the obscured population to which broad-line spectroscopic searches are insensitive. Two broad-line AGN exhibit markedly different mid-infrared emission properties, consistent with one being a little red dot (LRD) and the other either a typical AGN or an LRD with unusually strong hot-dust emission. The MIRI-selected AGN bolometric luminosity function at $z = 4.5$--$6$ yields number densities comparable to those of broad-line AGN and LRDs, suggesting that obscured AGN contribute significantly to the total AGN census at these epochs. The narrow-line AGN reside in diverse host environments, with evidence for both circumnuclear and host-galaxy-scale obscuration, pointing to multiple physical mechanisms at work. These results establish JWST/MIRI imaging as an indispensable component of a multi-faceted approach to a complete census of early supermassive black hole growth.

\end{abstract}


\section{Introduction} \label{sec:intro}

Supermassive black holes (SMBHs) reside at the centers of virtually all massive galaxies today \citep[e.g.][]{Magorrian98, Kormendy13} and are thought to play a fundamental role in shaping their host galaxies \citep[e.g.][]{Fabian12, Heckman14}. SMBHs are understood to grow primarily by accreting gas from the centers of their host galaxies, which are observed as active galactic nuclei \citep[AGN; e.g.][]{Antonucci93}. However, the physical origin of the earliest SMBHs, the environments in which they assembled their mass, and whether the co-evolutionary picture established at low redshift extends to early cosmic epochs all remain unresolved.

The discovery of hundreds of quasars at $z \gtrsim 5$ over the past decade (see review by \citealt{Fan23}) created a conundrum over the formation of SMBHs. These systems host SMBHs with masses of $\sim 10^{9-10} M_\odot$ when the universe is less than one billion years old \citep[e.g.][]{Wu15, Banados18}. Compounding this tension, observations of proximity zones created by the UV ionizing radiation of quasars at $z \sim 6$ revealed characteristically small sizes of these ionized regions, implying exceptionally short quasar lifetimes of $\sim 10^6$ yrs for $z \sim 6$ quasars, with some systems as short as $10^4$ yrs \citep[e.g.][]{Eilers18, Eilers21}. 
Quasar clustering measurements at $z \sim 6$ further support this picture, pointing to a short duty cycle of $<1$\%, and a similarly short lifetime of $\sim 10^6$ yrs \citep{Eilers24, Huang26}.  
Reconciling these short lifetimes with the observed SMBH masses would require either massive initial seeds far exceeding those expected from the remnants of the first stars \citep[e.g.][]{Loeb94, Bromm03, Lodato06}, or significant super-Eddington accretion \citep[e.g.][]{Volonteri05, Madau14, Volonteri15}, challenging standard models of black hole growth \citep[see review by][]{Inayoshi20}. One plausible explanation for this discrepancy is that the SMBHs experience the majority of their growth in a heavily dust-enshrouded phase \citep{Davies19, Satyavolu23}, during which the quasar's UV emission is suppressed, limiting its UV-bright duty cycle and the extent of its ionized proximity zone.

Indeed, observational evidence aligns with such a dust-obscured growth scenario. Deep X-ray surveys with Chandra and XMM-Newton \citep[e.g.][]{Iwasawa12, Vito18, Iwasawa20} indicate that the obscured fraction of AGN increases towards earlier epochs, rising from $\sim 20 \%$ at $z \sim 0$ to $\gtrsim 60\%$ at $z \sim 3$. Similarly, deep ALMA observations of massive galaxies show a rapid increase in the ISM column density towards high redshifts, implying that $80-90\%$ of the SMBHs at $z>6$ would be hidden from our view due to obscuration \citep{Gilli22}. Probing the full AGN population at these early epochs, including its obscured component, is therefore essential for a complete picture of early SMBH growth.

JWST has transformed our ability to select and characterize AGN at $z \gtrsim 5$. Spectroscopic observations using NIRSpec and the NIRCam wide-field slitless spectroscopy (WFSS) have revealed a remarkably abundant population of broad-line (BL) AGN at $z \gtrsim 5$ \citep[e.g.][]{Kocevski23, Larson23, Harikane23, Fujimoto24, Maiolino24, Matthee24, Taylor25}, with inferred number densities 10--100 times higher than those measured by the deepest X-ray surveys \citep{Maiolino24}. However, because these searches rely on broad Balmer emission lines, they are sensitive only to unobscured, Type I AGN and omit the obscured, Type II population. Given the high anticipated fraction of obscured AGN at these early epochs, BL spectroscopic searches likely represent only a fraction of the total AGN population at these epochs.

\begin{figure*}[t]
	\centering
		\includegraphics[width=0.99\textwidth]{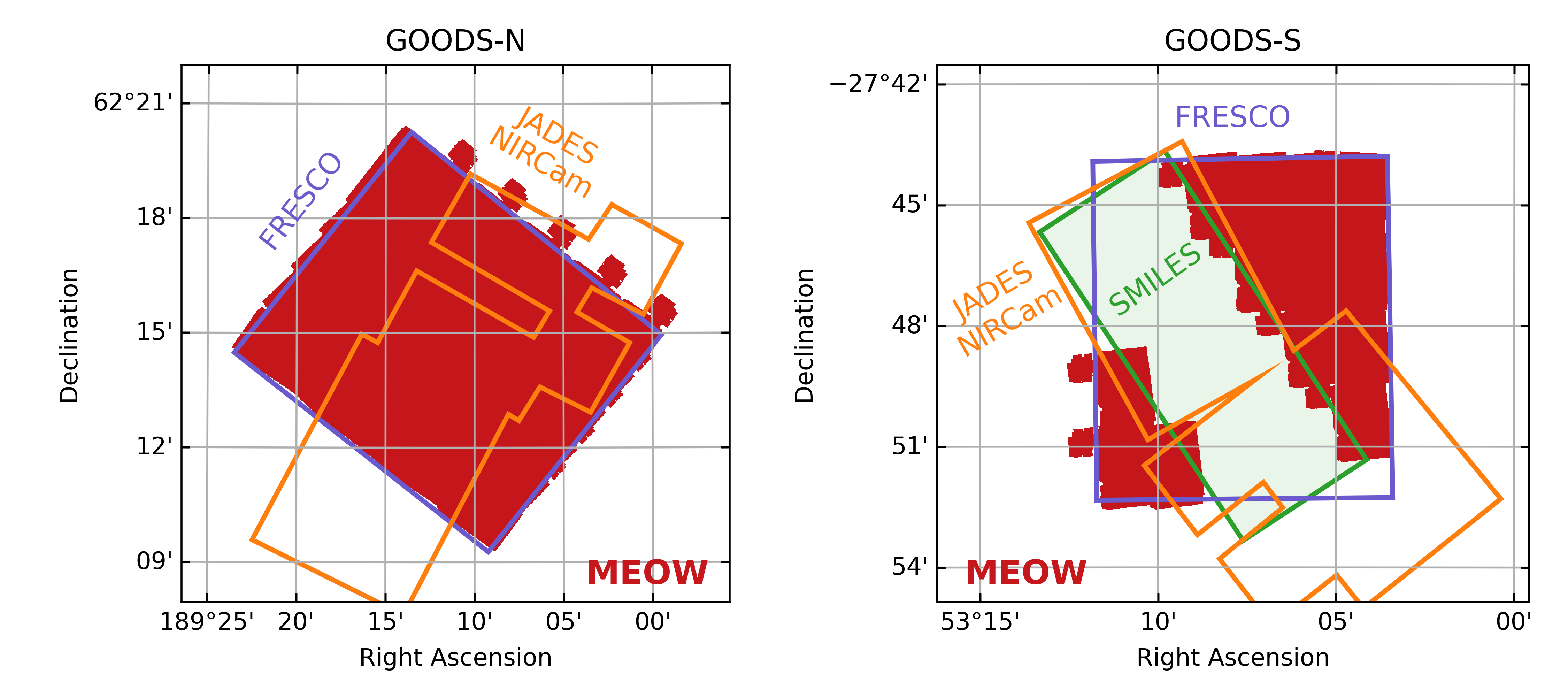}
		\caption{The MEOW survey footprint in the GOODS-N (left) and GOODS-S fields. The MEOW MIRI imaging (red shaded region) is designed to coincide with the NIRCam imaging and grism coverage of the FRESCO survey (blue outline). Parts of the footprint is additionally covered by NIRCam imaging from JADES (orange outline). Existing MIRI imaging from SMILES (green outline and shaded region) spans the central region of the GOODS-S field.}
		\label{fig:footprint}
\end{figure*}

While X-ray observations are regarded as a reliable tracer of AGN activity since X-ray photons are less susceptible to absorption, they suffer from incompleteness at high redshift ($z\gtrsim 3$). At these early epochs, a combination of abundant obscuring materials and greater cosmic distances reduces the detection rates to $\lesssim 30$--40\% \citep{Vito18}.
A powerful alternative for uncovering the obscured AGN population is to target the thermal emission of AGN-heated circumnuclear dust. The intense radiation field from the central engine raises dust grains to near their sublimation temperature at $\sim 1200$ K, producing a characteristic excess at rest-frame $\gtrsim 2 \ \um$ \citep{Barvainis87}. This method has proven successful in identifying obscured AGN up to $z \sim 3$ using near- to mid-infrared imaging from Spitzer/IRAC \citep[e.g.][]{Lacy04, Lacy13, Lacy15, Stern05, Donley12}. At higher redshifts, this hot-dust signature shifts into mid-infrared wavelengths at $\gtrsim 10 \ \um$, placing it beyond the reach of pre-JWST facilities except for hyper-luminous systems \citep[e.g.][]{Endsley22, Fujimoto22}.

The Mid-Infrared Instrument (MIRI) on JWST offers 10--100 times higher sensitivity and eight times better angular resolution than Spitzer, making it an ideal tool for detecting obscured AGN at high redshift through their hot dust emission. Early JWST studies leveraging MIRI have already yielded promising results \citep[e.g.][]{Yang23, Lyu24}, identifying AGN candidates out to $z \sim 6$ in small-area observations of the CEERS and SMILES fields, covering just 9--34 arcmin$^2$. To draw robust conclusions about the prevalence of obscured AGN at the highest redshifts, it is imperative to reach a statistical sample of $z>6$ sources, necessitating a significant expansion of the survey area for MIRI imaging at $\sim 10$--20 $\mu$m.

In this paper, we present MEOW: The MIRI Early Obscured-AGN Wide Survey, a wide-area MIRI imaging census of obscured AGN through cosmic time, with a focus on $z \gtrsim 5$. We describe the survey design in Section~\ref{sec:design} and the data reduction and ancillary data in Section~\ref{sec:data}. Section~\ref{sec:methods} details the galaxy and AGN selection procedures, and the resulting AGN sample at $z \gtrsim 5$ is presented in Section~\ref{sec:results}. We discuss the broader implications of our findings in Section~\ref{sec:discussion} and summarize our conclusions in Section~\ref{sec:conclusions}.

Throughout this paper, we assume a \citet{Planck20} cosmology of $H_0=67.4 ~ \mathrm{km~s}^{-1} ~\mathrm{Mpc}^{-1}$, $\Omega_\mathrm{m}=0.315$ and $\Omega_\mathrm{\Lambda} = 0.685$. All magnitudes are in the AB system.

\section{Survey Design}\label{sec:design}

MEOW (PID 5407; PI: G. Leung; Co-PIs: R. Endsley, S. Finkelstein) is a 74 hr JWST Cycle 3 MIRI imaging program. In this section, we describe the observational design of the MEOW survey.

\subsection{Target Fields}

MEOW targets the central regions of the GOODS-N and GOODS-S fields \citep{Giavalisco04} to exploit the wealth of multiwavelength legacy observations therein. Our survey footprint is designed to cover the First Reionization Epoch Spectroscopically Complete
Observations (FRESCO; PID 1895; \citealt{Oesch23}), which obtained grism spectra with NIRCam WFSS in the F444W filter, and imaging observations in the F182M, F210M and F444W filters. The F444W grism spectra from the FRESCO survey detect [SIII] at $z=2.5-4.5$, H$\alpha$+[NII] at $z=5-6.5$ and [OIII]+H$\beta$ up to $z=7.9$, providing robust spectroscopic redshifts for our MIRI-detected sources. In addition, GOODS-N was also targeted by the Complete NIRCam Grism Redshift Survey (CONGRESS; PID 3577; PIs: E. Egami, F. Sun), which extended the grism coverage to F356W, enhancing spectroscopic redshift coverage at $z=3.8-5$. 
The GOODS fields are also home to rich legacy HST imaging data. The GOODS \citep{Giavalisco04} and CANDELS \citep{Grogin11, Koekemoer11} programs provide deep UV, optical and near-IR imaging throughout the MEOW footprint, offering the photometric coverage from 0.4 to 1.6 \um . The combination of JWST FRESCO and HST CANDELS imaging ensures continuous photometric coverage from 0.4 to 4.4 \um \ over the full survey footprint.

In addition, the majority ($\sim 70 \%$) of the MEOW footprints in the GOODS fields are covered by extensive NIRCam imaging from the JWST Advanced Deep Extragalactic Survey (JADES; \citealt{Eisenstein26}). The JADES NIRCam data covers up to 13 filters in 0.9--4.4 \um \ down to 28--30 mag, providing excellent sampling of the SED and allowing exquisite photometric redshift measurements and estimation of physical parameters. Additionally, our footprint in the GOODS-S field overlaps with the JWST Extragalactic Medium-band Survey (JEMS; PID 1963;  \citealt{Williams23}).

\subsection{MIRI Observations}

We use a total of 43 MIRI pointings to cover the FRESCO footprint in the GOODS-N and GOODS-S fields. The GOODS-N field is observed by 30 pointings producing a $5\times 6$ mosaic. Parts of the GOODS-S field are targeted with MIRI imaging by the Systematic Mid-infrared Instrument Legacy
Extragalactic Survey (SMILES; PID 1207; \citealt{Rieke24}) in all eight MIRI filters. We observe the GOODS-S field with 13 pointings to cover the regions not observed by SMILES. This results in a final survey area of $95$ arcmin$^2$ in the two fields combined. We show our survey footprint in Figure \ref{fig:footprint}.

We observe in the F1000W and F2100W filters, selected to detect the AGN-heated hot dust emission feature at high redshift. The F2100W filter captures the AGN infrared signature at rest-frame 2--3 \um \ up to $z \sim 8$,  while the F1000W filter constrains the slope of the IR continuum, which is crucial for distinguishing an AGN IR `bump' from a smoothly rising spectrum in dusty star-forming galaxies. We adopt the CYCLING-LARGE dither pattern, well suited to mitigate cosmic ray showers. Using the FASTR1 readout pattern, we reach an exposure time of 12 and 51 minutes per pointing in the F1000W and F2100W filters, respectively. For comparison, the exposure times of SMILES in the same filters are 11 and 36 minutes. Table \ref{tab:phot} outlines the exposure setup in each filter.

\begin{deluxetable*}{lccccccccc}
\tablecaption{Imaging and Photometry Parameters}
\tablehead{
\colhead{Filter} & \colhead{$N_\mathrm{groups}$} & \colhead{$N_\mathrm{int}$} & \colhead{$N_\mathrm{exp}$} & \colhead{Exp. Time} &
\colhead{Aperture} & \colhead{Aperture} & \colhead{$5\sigma$ Depth} & \colhead{50\% Compl.} & \colhead{80\% Compl.}
\\
\colhead{} & \colhead{} & \colhead{} & \colhead{} & \colhead{(s)} &
\colhead{Radius} & \colhead{Correction} & \colhead{(\uJy)} & \colhead{(\uJy)} & \colhead{(\uJy)}
\\
\colhead{(1)} & \colhead{(2)} & \colhead{(3)} & \colhead{(4)} & \colhead{(5)} & \colhead{(6)} & \colhead{(7)} & \colhead{(8)} & \colhead{(9)} & \colhead{(10)}
}
\startdata
F1000W & 65 & 1 & 4 & 722 & $0\farcs3$ & 1.7348 & 0.499 & 0.335 & 0.429 \\
F2100W & 30 & 9 & 4 & 3086 & $0\farcs5$ & 1.8288 & 3.59 & 2.90 & 3.66 \\
\enddata
\tablecomments{Column (1): MIRI filter. Column (2): number of groups. Column (3): number of integrations. Column (4): number of dithers. Column (5): total exposure time over the dithers. Column (6): radius of circular apertures used for photometry. Column (7): aperture correction to total flux. Column (8): $5\sigma$ photometric depths measured by randomly placed apertures with radii equal to column (7). Columns (9) and (10): 50\% and 80\% completeness measured by source injection and recovery.}
\end{deluxetable*}\label{tab:phot}

\section{Data}\label{sec:data}

In this section, we describe the reduction and photometry of the MIRI data, as well as the ancillary data in the GOODS fields used in our analysis.

\subsection{MIRI Data}

For our reduction and scientific analysis, we combine MEOW data with publicly available MIRI imaging datasets in the F1000W and F2100W filters in the GOODS fields for completeness. These primarily include data from SMILES (PID 1207) in GOODS-S, as well as smaller datasets from PIDs 2926, 4762 in GOODS-N, and PIDs 1283, 4498 and 6511 in GOODS-S.

\subsubsection{Image Reduction}

The combined MIRI dataset was processed through the JWST Rainbow pipeline, adopting the procedures detailed in \cite{Perez-Gonzalez24}. A complete description of the reduction, including efforts in other legacy fields such as A2744, COSMOS, EGS, and UDS is deferred to Pérez-González et al. (in prep., see also Magee et al., in prep.). We provide a brief overview of the methodology here. The JWST Rainbow pipeline builds upon the official JWST reduction framework (version v1.19.41, calibration context \texttt{jwst\_1413.pmap}), augmenting it with tailored processing steps designed to correct for background inhomogeneities inherent to MIRI imaging. An important feature of this approach is the ``superbackground'' technique \citep{Perez-Gonzalez24, Yang23a}, whereby background maps are derived for each individual exposure by drawing upon the complete set of images acquired throughout the observing campaign, with cataloged sources masked to prevent flux contamination. The resulting background is highly uniform in both its absolute level and associated noise. In quantitative terms, the method suppresses the standard deviation of background pixels in individual cal.fits files by a factor of $\sim$3.5 compared to the default pipeline, translating to a 0.8 mag gain in the depth of the final mosaic. Further technical details are provided in Appendix A of \citet{Perez-Gonzalez24}, while analogous noise improvements have been demonstrated for the same reduction technique but different approaches to estimate depths \citep[e.g.][]{Yang23a, Alberts24, Ostlin25, Backhaus25}. The final mosaics are resampled to a pixel scale of 0\farcs06.

\subsubsection{Source Detection}

We use \texttt{SExtractor} (\citealt{Bertin96}, v2.28.2) to perform source detection in each filter independently with the \texttt{MAP\_RMS} weighting mode. For the RMS map, we take the pipeline ERR map, and rescale it by 1.48 times the median absolute deviation of the science image pixel values, so that the ERR map values represent the empirical $1\sigma$ fluctuations in the science image, which could capture systematic noise not represented by the pipeline ERR map. In practice, this is equivalent to adopting a \texttt{DETECT\_THRESH} scaled by a constant, but the rescaling facilitates a more uniform interpretation of the adopted thresholds across different images.

We adopt a hot+cold scheme to detect faint objects while avoiding over-deblending of bright, extended objects \citep[e.g.][]{galametz13}. We first run a cold detection process to detect bright, extended sources with more conservative \texttt{DETECT\_THRESH}, \texttt{MINAREA}, and \texttt{DEBLEND\_MINCONT} parameters, followed by a hot run for faint sources with more aggressive detection and deblending. The cold and hot catalogs are then combined as follows. First, all sources in the cold catalog are included. Then for the hot catalog, sources are only included if their source position falls outside of the segmentation map of the cold catalog, dilated by 6 and 8 pixels in F1000W and F2100W, respectively. Finally, we remove spurious detections from noise spikes that show significantly narrower brightness profiles than the PSF of the filter by measuring the fraction of flux enclosed in a very small aperture. For F1000W, we remove sources where the ratio of flux enclosed in a 1-pixel (60 mas) radius aperture is $>40\%$ of that in a 4-pixel radius aperture. For F2100W, we use the same fraction, but for 2- and 8-pixel apertures due to the larger PSF in the filter. Visual inspection of the removed sources confirms that they are primarily noise spikes in regions with shallower exposure time and/or near image edges.

\subsubsection{Photometry}

We measure fluxes using circular apertures centered at the source position with radii containing $\approx 60\%$ of the flux of the PSF in the corresponding filter, adopting $0\farcs3$ and $0\farcs5$ radius apertures for F1000W and F2100W, respectively. We apply aperture correction to the fluxes based on the MIRI instrument team flux calibration results \citep{gordon25}, which are available in the reference file \texttt{jwst\_miri\_apcorr\_0014.fits}. This results in an aperture correction of 1.73 and 1.83 for the F1000W and F2100W filters, respectively. Finally, we correct for Galactic extinction using the \citet{cardelli89} Milky Way attenuation curve and assuming an $E(B-V)$ of 0.0103 and 0.0069 for GOODS-N and GOODS-S, respectively \citep{schlafly11}.

\subsubsection{Photometric Uncertainties and Depths}

We determine the photometric uncertainties empirically to account for correlated noise. For each mosaic, we randomly place 5,000 non-overlapping apertures, with the same radii as those used for flux measurements, avoiding invalid pixels in the science image or source pixels with non-zero values in the segmentation map. To measure the $1\sigma$ flux uncertainty, we calculate the median absolute deviation of the fluxes within the random apertures, and apply a factor of 1.48 to convert to a Gaussian-like standard deviation. In order to account for varying depths across the mosaic, we scale this $1\sigma$ uncertainty by the ratio between the value of the error array at the source center and the median of the entire error array. Finally, we apply the same aperture correction to the flux uncertainty as the flux to correct to the total. We measure the $5\sigma$ depth by taking the aperture-corrected $1\sigma$ flux uncertainty without scaling with the error map values, and multiplying it by 5. The $5\sigma$ photometric depths are reported in Table \ref{tab:phot}.

\subsubsection{Detection Completeness}
\label{sec:det_comp}

\begin{figure}[!t]
	\centering
        \includegraphics[width=0.45\textwidth]{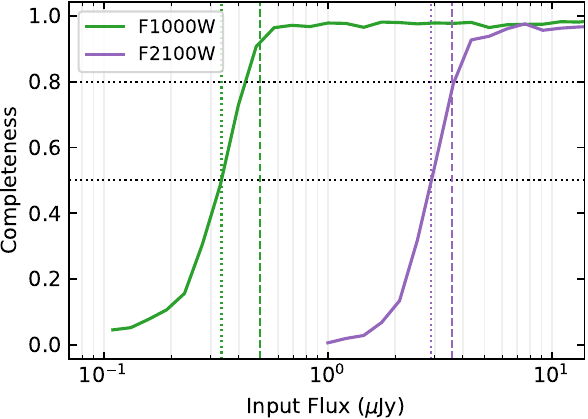}
		\caption{Detection completeness as a function of flux for the F1000W (green) and F2100W (purple) filters, measured from the simulation and recovery of point sources in the science mosaics. The vertical dotted and dashed lines indicate the 50\% completeness limit and $5\sigma$ depth for each filter, respectively.}
		\label{fig:completeness}
\end{figure}

We estimate our point-source detection completeness by injecting and recovering mock sources in the science images. To produce the mock sources, we select compact sources from our detections based on their location in the size-magnitude space, and stack them to obtain a model source profile. We then generate mock sources with magnitudes in a log-normal distribution, and randomly place them in the science images. We do not avoid known source pixels, as source blending due to projection effects can also occur in actual astrophysical objects. We inject $10^3$ sources per iteration, with 10 iterations per image, achieving a total of $10^4$ injected sources. The images are then processed through the same source detection procedure as the real data. The completeness is measured as the ratio of mock sources recovered to injected as a function of the input magnitude. We show the completeness is Figure \ref{fig:completeness}. We find $80\%$ completeness of 0.43 and 3.66 \uJy \ in the F1000W and F2100W filters, respectively. These $80\%$ completeness values are comparable to the $5\sigma$ depths reported in Table \ref{tab:phot}.

\subsection{Ancillary Data}

\subsubsection{JWST/NIRCam and HST Imaging} \label{sec:nircam}

For the existing NIRCam and HST imaging in the GOODS fields, we use photometric catalogs from the UNICORN (Uniform Near-Infrared CatalOgs
from Robust imagiNg) project (S. L. Finkelstein et al. in prep.), which provides photometric catalogs over major JWST legacy fields. The photometric procedures are largely based on those described in \citet{Finkelstein24}, which performs PSF-matched photometry with various aperture corrections to obtain accurate colors, total flux and flux uncertainties. The catalog uses a F277W+F356W composite as the primary detection band, supplemented by F444W to include areas not covered by the first two filters (these regions \emph{are} covered by {\it HST}/WFC3 imaging, included in this catalog).  As the MEOW footprint is designed around that of the FRESCO survey, our survey area is fully covered by F444W imaging, with the exception of very small gaps near the detector edges. 

\subsubsection{NIRCam WFSS}
The FRESCO data were reduced following the Simulation Based Extraction (SBE) described in \citet{Pirzkal17}. We used the contemporaneous F444W imaging data to drive the simulations and extraction process. The WFSS configuration for the NIRCam grisms were from CRDS jwst\_1126.pmap. Details of the background subtraction, contamination estimation, and optimally weighted extraction processed can be found in \citet{Pirzkal17}.

\subsubsection{NIRSpec Spectroscopy}

We make use of publicly available NIRSpec data products from the JADES Data Release 4 \citep{Eisenstein26, Curtis-Lake25, Scholtz25}. In particular, we take spectroscopic redshifts from the spectroscopic catalog, as well as reduced 1D spectra for our analysis. Readers are also referred to \citet{Bunker24} and \citet{D'Eugenio25} for additional details on JADES spectroscopic data reduction.

\subsubsection{Submillimeter Data}

Both GOODS fields are covered by rich submillimeter data. For this work, we make use of published source catalogs from the SUPER GOODS survey, a deep SCUBA-2 survey of both of the GOODS fields. We use 450 \um \ fluxes from the catalogs of \citet{Barger22} and 850 \um \ fluxes from \citet{Cowie17} and \citet{Cowie18}. The SCUBA-2 data from SUPER GOODS reach a 450 \um \ RMS of 1.14 and 1.86 mJy in GOODS-N and GOODS-S, and near the confusion limit of 1.65 mJy in the central parts of the fields.

\subsection{Photometric Redshift}

We measure photometric redshifts for all the sources in the UNICORN catalog using \texttt{Lazy} \footnote{\url{https://github.com/hollisakins/Lazy.jl}}, which is a Julia-based version of the template-based fitting code \texttt{EAZY} \citep{Brammer08}. Similar to \citet{Leung23} and \citet{Finkelstein24}, we use the 12 ``\texttt{tweak\_fsps\_QSF\_12\_v3}'' templates, supplemented by six additional templates of \citet{Larson23b}, which are optimized for the blue colors of very high-redshift galaxies. We adopt a non-informative prior on luminosity and include a systematic error of 5\% of the observed flux. The fitting is performed with all the available JWST/NIRCam and HST filters in the UNICORN catalog. The MIRI photometry is not used in the photometric redshift fits, as the galaxy templates employed do not generally account for AGN or dust emission, which is expected to dominate the observed MIRI bands.

\begin{deluxetable*}{llcc}
\tablecaption{\textsc{cigale} Model Parameters}
\tablehead{
\colhead{Module} & \colhead{Parameter} & \colhead{Symbol} & \colhead{Values} 
}
\startdata
\multirow{2}{4.5cm}{Star formation history \\ \texttt{sfhdelayed}} & Stellar e-folding time & $\tau_\mathrm{star}$ & 0.5, 1, 2, 5 Gyr \\
& Stellar age & $t_\mathrm{star}$ & 0.1, 0.2, 0.5, 1, 2, 5 Gyr \\
\hline
\multirow{2}{4.5cm}{Simple stellar population \\ \texttt{bc03}} & Initial mass function & \texttt{imf} & Chabrier \\
& Metallicity & $Z$ & 0.02 \\
\hline
\multirow{2}{4.5cm}{Nebular emission \\ \texttt{nebular}} & Ionization parameter & $\log U$ & $-2.0$ \\
& Gas metallicity & $Z_\mathrm{gas}$ & 0.02 \\
\hline
\multirow{3}{4.5cm}{Dust attenuation \\ \texttt{dustatt\_modified\_starburst}} & \multirow{2}{5cm}{Nebular line color excess} & \multirow{2}{2cm}{$E(B-V)_\mathrm{line}$} & 0, 0.02, 0.05,  \\
& & & 0.1--1.5 (0.1 interval)\\
& Line to ratio $E(B-V)$ ratio & $\frac{E(B-V)_\mathrm{line}}{E(B-V)_\mathrm{cont}}$ & 1 \\
\hline
\multirow{4}{4.5cm}{Galactic dust emission \\ \texttt{dl2014}} & PAH mass fraction & $q_{\rm PAH}$ & 0.47, 2.5, 7.32 \\
& Minimum radiation field &  $U_{\rm min}$ & 0.1, 1.0, 10, 50 \\
& \multirow{2}{5cm}{Fraction of PDR emission} &  \multirow{2}{1cm}{$\gamma$} & 0.01, 0.02, 0.05, 0.1,  \\
& & & 0.2, 0.5, 0.9 \\
\hline
\multirow{5}{4.5cm}{AGN \\ \texttt{skirtor2016}} & Average edge-on optical depth at $9.7 \mu$m & $\tau_{9.7}$ & 3, 5, 7, 9, 11 \\
& Viewing angle & $\theta_{\rm AGN}$ & 70$^\circ$ \\
& \multirow{2}{5cm}{AGN contribution to IR luminosity} &  \multirow{2}{1cm}{\fagn} & 0.0, 0.05, 0.1, 0.2, 0.3, 0.4,  \\
&  &  & 0.5, 0.6, 0.7, 0.8, 0.9, 0.99 \\
& Rest-frame wavelength where \fagn \ is defined & $\lambda_{\rm AGN}$ & 3--30 \um \\
\enddata
\end{deluxetable*}\label{tab:cigale}

\section{AGN Identification Methodology} \label{sec:methods}

In this section, we describe our procedures to select AGN candidates at $z \gtrsim 5$. For this practice, we first select a sample of sources detected independently in the F1000W and F2100W filters, as well as the NIRCam detection bands. To do this, we cross-match the F1000W and F2100W catalogs to the NIRCam detections (described in Section \ref{sec:nircam}), defining sources with a match within $0\farcs5$ as a good match. This results in a total of 1518 sources in GOODS-N, and 1191 sources in GOODS-S. 

\subsection{Selection of $z \gtrsim$ 5 Galaxies} \label{sec:gal}

We first select galaxy candidates at $z \gtrsim 5$ using a combination of photometric and spectroscopic selection criteria.

\subsubsection{Photometric Selection}

We select a sample of $z \gtrsim 5$ galaxy candidates using selection criteria based on a combination of quantities derived from photometric redshift fitting. We denote the probability density function of the photometric redshift as $P(z)$. For each integer redshift $z$, we integrate $P(z)$ in the interval $z\pm 0.5$. We denote the integer redshift with the highest such integral as $S_z$. Our selection criteria are as follows:
\begin{enumerate}
    \item Best-fit photometric redshift $(z_a) > 4.5$.
    \item $\Delta \chi^2 > 4$ calculated as the difference between the best-fit $\chi^2$ at below and above $z=4$, corresponding to a $2\sigma$ significance.
    \item $\int P(z>4) > 0.9$, implying an over 90\% probability that the redshift is greater than 4.
    \item $S_z \ge 5$ 
\end{enumerate}

These criteria result in 16 candidates in GOODS-N and 8 in GOODS-S. This initial sample is then vetted by visually inspecting the image cutouts in the NIRCam and MIRI bands. This resulted in four sources in each field being removed from the sample. Among these eight removed sources, seven are excluded based on their NIRCam images, with four located on a diffraction spike of a nearby star, one blended with a bright extended neighbor, and two highly extended favoring a $z < 4$ secondary photometric redshift solution. The remaining excluded source is located near the edge of the MIRI images with elevated noise potentially leading to spurious detection. This results in a sample of 16 $z \gtrsim 5$ photometrically selected candidates, with 12 and 4 in GOODS-N and GOODS-S, respectively. Three of the candidates are at $z > 7$.

\subsubsection{Spectroscopic Selection}
We supplement our selection using spectroscopic redshifts from the FRESCO survey by cross-matching the MIRI and NIRCam detected sources to the FRESCO team H$\alpha$ \citep{Meyer24} and [OIII] \citep{Covelo-Paz25} emitter catalogs, covering galaxies at $z=4.9-6.7$ and $z=6.8-9.0$, respectively. Among the photometrically-selected galaxies, we find spectroscopic redshift matches for all three of the $z>7$ galaxies, and four of the 13 $z\approx 5-7$ galaxies. No additional MIRI-detected galaxies are found in the H$\alpha$ or [OIII] emitter catalogs, leaving the final sample at 16 objects.

We then measure spectroscopic redshifts for these 16 objects using the extracted FRESCO grism spectra and NIRSpec grating spectra from JADES, where available. The grism spectra confirm spectroscopic redshifts for all seven of the cross-matched objects, and yield two additional redshifts at $z\sim 5$ via H$\alpha$ detection. The NIRSpec spectra yield a spectroscopic redshift via H$\alpha$ detection for one $z=4.7$ galaxy. Additionally, two objects within the photometrically-selected sample have published spectroscopic redshifts. GOODS-N 6404 has a spectroscopic redshift at $z=5.182$, identified through NOEMA observations of its CO emission line \citep{Lagache26}, while GOODS-S 18381 has a spectroscopic redshift of $z=4.99$ reported in the CANDELS catalog \citep{Kodra23}. However, no emission lines are detected in the grism spectra for these two objects.  

\begin{figure}[!t]
	\centering
		\includegraphics[width=0.45\textwidth]{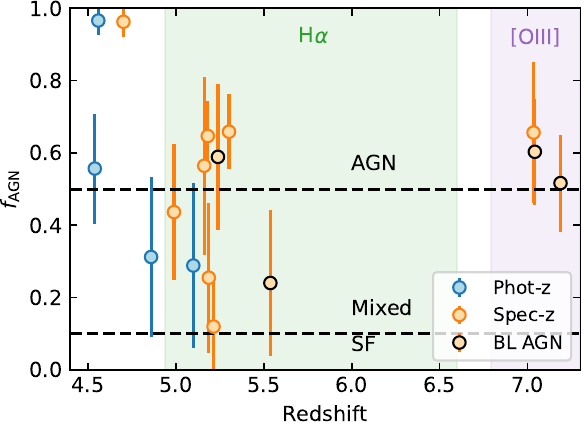}
		\caption{The distribution of \fagn \ and redshift for the AGN sample at $z \gtrsim 5$. The orange and blue points denote objects where spectroscopic and photometric redshifts, respectively, are available. Objects known to be BL AGN in prior studies are denoted with black outlines. The redshift ranges where H$\alpha$ and [OIII] are covered by the FRESCO F444W grism spectra are shown by the green and purple shaded regions, respectively. Spectroscopic redshifts are available in 11 out of 12 sources within the FRESCO spectral coverage. Out of the 16 AGN in our sample, 12 are new AGN not previously identified by BL selection.}
		\label{fig:sample}
\end{figure}

This results in a total of 12 spectroscopic redshifts out of the 16 objects in our sample. Among the four objects without a spectroscopic redshift, three of them have a best-fit photometric redshift of $z=4.5-4.9$, where the H$\alpha$ emission line is expected to fall out of the F444W filter, where grism spectra from FRESCO are available. Conversely, a spectroscopic redshift is obtained in eight out of the nine objects with a best-fit photometric redshift within the H$\alpha$ coverage in F444W, indicating that our photometric selection is robust. We show the redshift distribution of the full sample in Figure \ref{fig:sample}.

\subsection{Selection of $z \gtrsim 5$ AGN}\label{sec:agn}

In this subsection, we describe the procedures to identify AGN within the photometrically and spectroscopically selected $z\gtrsim 5$ galaxy sample.

\begin{figure*}[!t]
	\centering
		\includegraphics[width=0.85\textwidth]{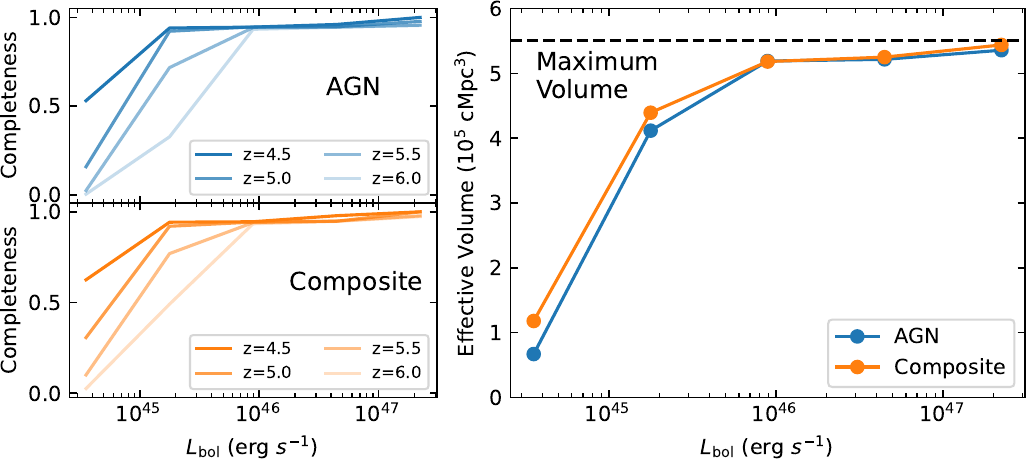}
		\caption{Left: The detection completeness as a function of AGN bolometric luminosity for the AGN (top) and composite (bottom) sources at $z=4.5, 5.0, 5.5$ and 6.5. Right: The effective volume for the AGN (blue) and composite (orange) sources in the $z=4.5$--6 redshift bin. The black dashed line shows the maximum volume. The final effective volume is the average of the AGN and composite volume weighted by the number of sources in each classification in each luminosity bin.}
		\label{fig:volume}
\end{figure*}

\subsubsection{SED Modeling}
To quantify the AGN emission in the $z \gtrsim 5$ galaxies and identify AGN, we model their SED using \texttt{CIGALE} \citep[][v2022.1]{Boquien19, Yang20, Yang22}.
We fix the redshift to the spectroscopic redshift when available. Otherwise, we adopted the best-fit photometric redshift from \texttt{EAZY}. SED fitting is performed using the multi-band photometry described in Section \ref{sec:data}, spanning 20 (22) optical to IR filters with HST and JWST in GOODS-N (S), as well as two submillimeter bands with SCUBA-2. For sources that are undetected by SCUBA-2, we use upper limits based on the median RMS for the 450 $\mu$m band and the confusion limit for the 850 $\mu$m band.

Two of our $z \gtrsim 5$ galaxies are in the sample of broad H$\alpha$ emitters reported in \citet{Matthee24}. As the broad H$\alpha$ emission could significantly boost the broadband photometry in the F444W filter, and \texttt{CIGALE} does not model broad Balmer emission lines from type 1 AGN, we subtract the broad line fluxes from the photometry for these objects. The corrected flux is calculated by $F_\lambda^c = F_\lambda - F_\mathrm{line} \Delta \lambda ^{-1}$, where $F_\lambda$ is the uncorrected broadband flux density, $F_\mathrm{line}$ is the integrated broad H$\alpha$ line flux, and $\Delta \lambda$ is the effective width of the F444W filter. We then convert $F_\lambda$ to $F_\nu$ using the pivot wavelength of the F444W filter.

In our SED model, we include the starlight and nebular emission from the stellar population, reprocessed galactic dust emission due to the attenuated starlight, and emission from the AGN including the accretion disk and dusty torus. We summarize the \texttt{CIGALE} fitting parameters in Table \ref{tab:cigale}, and highlight the model setup below.

\begin{figure*}[!t]
	\centering
		\includegraphics[width=0.9\textwidth]{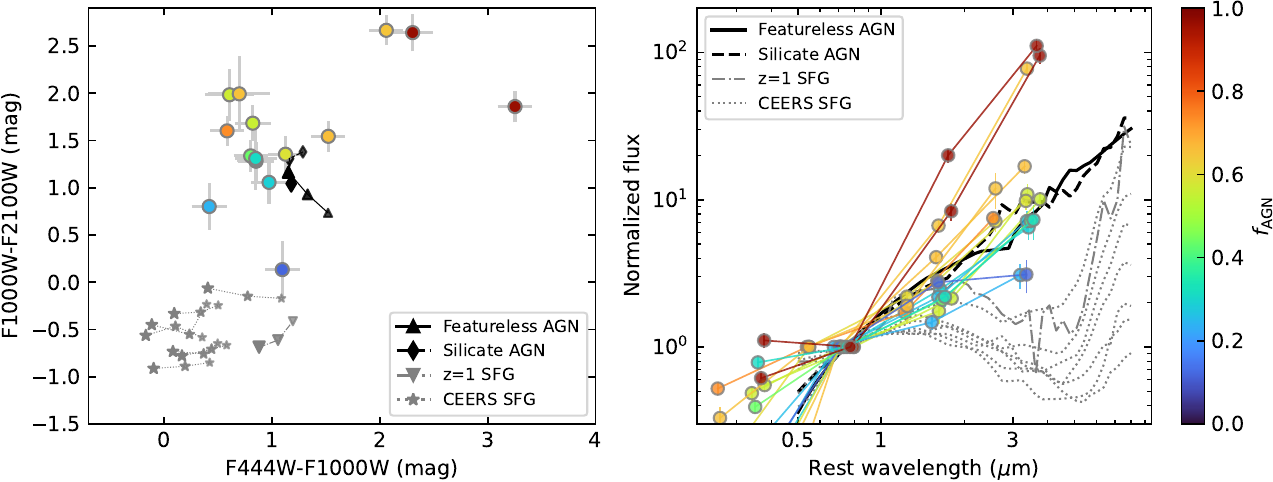}
		\caption{Left: The F1000W-F2100W vs. F444W-F1000W color-color diagram for the $z\gtrsim 5$ AGN sample. Each object is color-coded by their best-fit \fagn . We overplot the evolutionary tracks of AGN (black) and star-forming galaxy (gray) templates from \citet{Kirkpatrick12} and \citet{Kirkpatrick23} at $z=4, 5$ and 6. Right: The observed rest-frame SED of the AGN sample, color-coded by the best-fit \fagn \ of each object. The AGN and star-forming galaxy templates are shown by the black and gray lines, respectively. The \fagn \ value provides a strong proxy for separating AGN and galaxy SED.}
		\label{fig:color}
\end{figure*}

For the stellar emission, we adopt a standard delayed-$\tau$ star formation history using the \citet{Bruzual03} stellar population synthesis model. We assume a \citet{Chabrier03} initial mass function. Nebular emission from HII regions is modeled with the prescription of \citet{Villa-Velez21}. For dust attenuation of the galaxy emission, we employ the \texttt{dustatt\_modified\_starburst} module in \texttt{CIGALE}, which is a \citet{Calzetti00} attenuation law extended below 150 nm with the \citet{Leitherer02} curve. To accommodate potential dusty galaxies in our sample and distinguish them from obscured AGN, we allow $E(B-V)$ values from 0 to 1.5.

The galactic dust emission is modeled using the templates of \citet{Draine14}, which include a diffuse dust component and a photodissociation region component associated with star formation. The mass fraction of polycyclic aromatic hydrocarbons (PAHs) relative to the total dust mass is a free parameter in the model. The total dust luminosity is equal to the attenuated luminosity of the starlight, following the principle of energy balance in \texttt{CIGALE}.

We model the AGN emission using the \texttt{skirtor} clumpy torus model of \citet{Stalevski12, Stalevski16}. A key parameter in the model is the relative strength between the AGN and galaxy emission, set by the \fagn \ parameter. We allow \fagn \ to vary between 0 and 0.99, with \fagn \ measured in the rest-frame wavelength range of $3-30$ \um . Following \citet{Yang23}, we only allow viewing angles corresponding to type 2 AGN. Type 1 models are highly degenerate with the stellar emission in the rest-frame UV and optical, and type 1 AGN are generally rare compared to their type 2 counterparts within IR-selected samples. Importantly, the bolometric luminosity of the AGN, the key derived quantity, is strongly constrained by the hot dust emission from the torus, which is largely insensitive to the viewing angle and falls directly within the MIRI bandpass at our target redshifts. 

GOODS-N 107919 in our sample corresponds to the $z=7.2$ red quasar GNz7q \citep{Fujimoto22, Fei26}. Detailed SED modeling and analysis of this source using data including MEOW MIRI photometry is presented in \citet{Fei26}. For consistency, we adopt the \texttt{CIGALE} model configurations in \citet{Fei26} for this object.

\subsubsection{Sample Completeness}\label{sec:samp_comp}

We now estimate the completeness of the AGN sample. For an object to be selected, it has to be detected in NIRCam and both MIRI filters. Given the significantly deeper imaging in NIRCam, we assume that the NIRCam detection is complete within the magnitude range of our sample, and the sample completeness is dominated by the completeness in MIRI. For instance, the object in our sample with the faintest F444W flux is at 26.2 mag, significantly brighter than the FRESCO $5\sigma$ depth of 28.2 mag \citep{Oesch23}.

The detection completeness in MIRI as a function of flux is presented in Section \ref{sec:det_comp}. We translate this to a function of \lbol \ and redshift as follows. We first construct empirical SED templates by median stacking the model SEDs of objects in our sample. We do this in two bins in \fagn , for $\fagn > 0.5$ and $0.1 < \fagn < 0.5$, corresponding to AGN and composite sources, respectively (see Section \ref{sec:sample}). We then derive an empirical relation between \lbol \ and the monochromatic luminosity at rest-frame 3 \um \ ($L_\mathrm{3\um}$) from our SED model for each object. Finally, we scale and shift the median template in each \fagn \ bin using the \lbol \ correlation in a grid of \lbol \ and redshifts, and calculate the observed fluxes in the F1000W and F2100W filters. The final detection completeness is the product of the detection completeness in the F1000W and F2100W filters at the corresponding fluxes for the \lbol \ and redshift in question. We show the completeness for the AGN and composite sources as a function of \lbol \ at $z=4.5-6$ in the left panels of Figure \ref{fig:volume}.

\section{AGN Identification Results}\label{sec:results}

In this section, we describe the results of the SED modeling and AGN identification at $z \gtrsim 5$. 

\subsection{AGN Sample at $z \gtrsim$ 5}\label{sec:sample}

We identify AGN in the sample based on the \fagn \ parameter from the SED modeling results, adopting the Bayesian output from \texttt{CIGALE} as our fiducial values. To examine the separation of AGN and galaxies by \fagn , we show in the left panel of Figure \ref{fig:color} the NIRCam-MIRI colors for our sample, color coded by \fagn . There is a strong correlation between \fagn \ and the observed colors, with increasing \fagn \  at redder F1000W-F2100W colors, and a small fraction of objects with the highest \fagn \ occupying red colors in both F1000W-F2100W and F444W-F1000W.

\begin{figure*}[!t]
	\centering
        \includegraphics[width=0.49\textwidth]{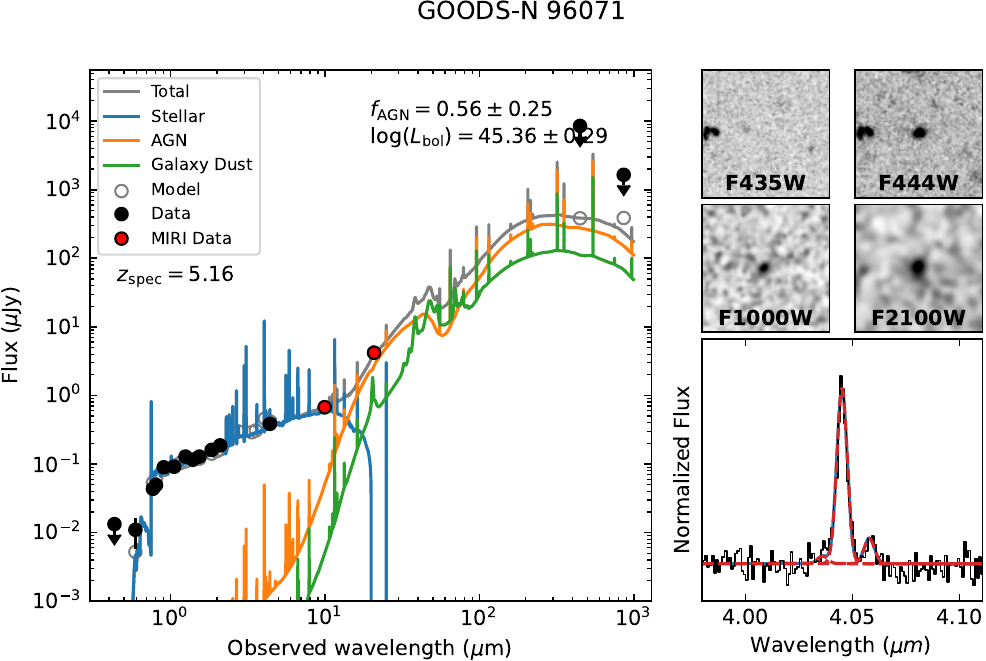}
        \hfill
        \includegraphics[width=0.49\textwidth]{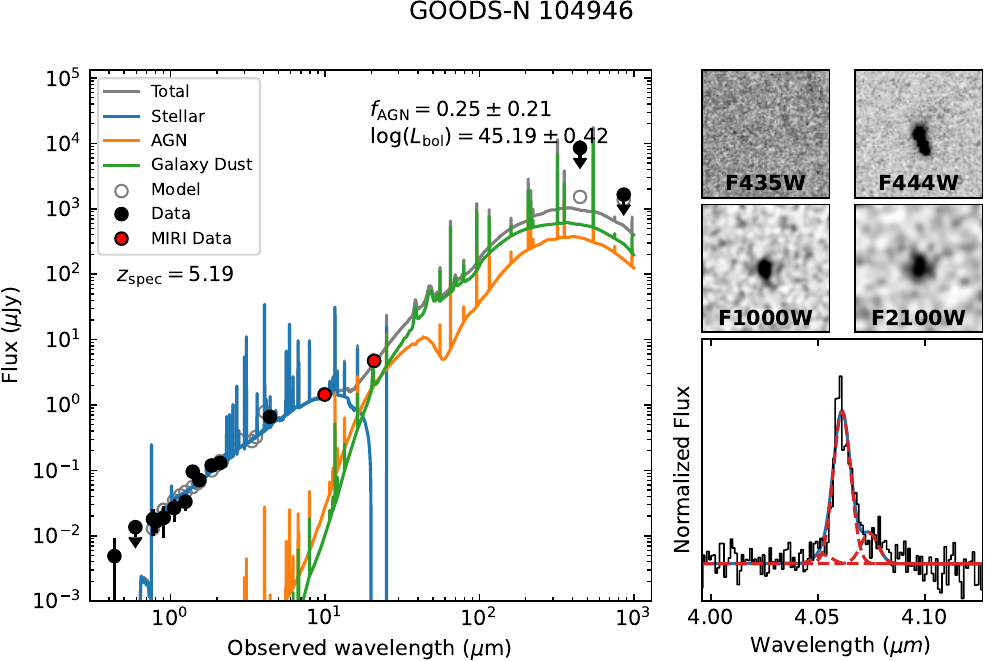}
        \\
	        \includegraphics[width=0.49\textwidth]{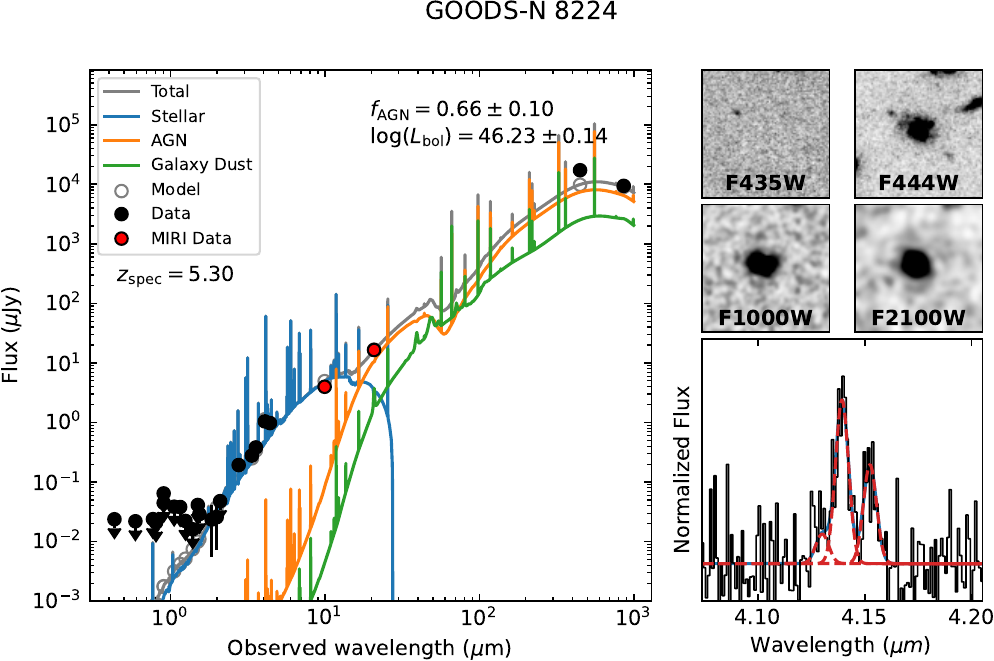}
        \hfill
        \includegraphics[width=0.49\textwidth]{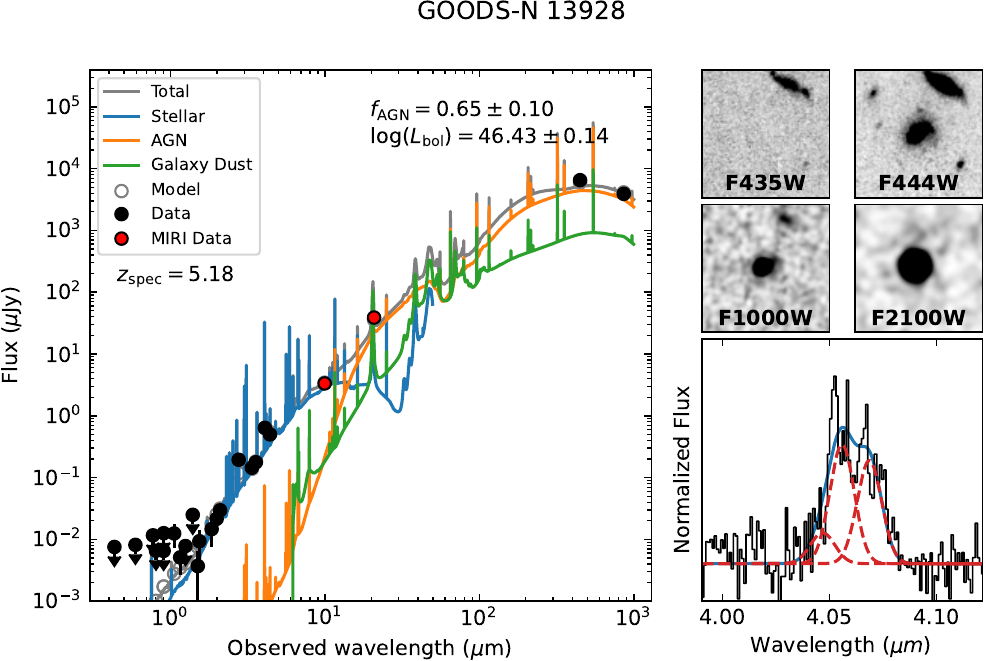}
        \\
        \includegraphics[width=0.49\textwidth]{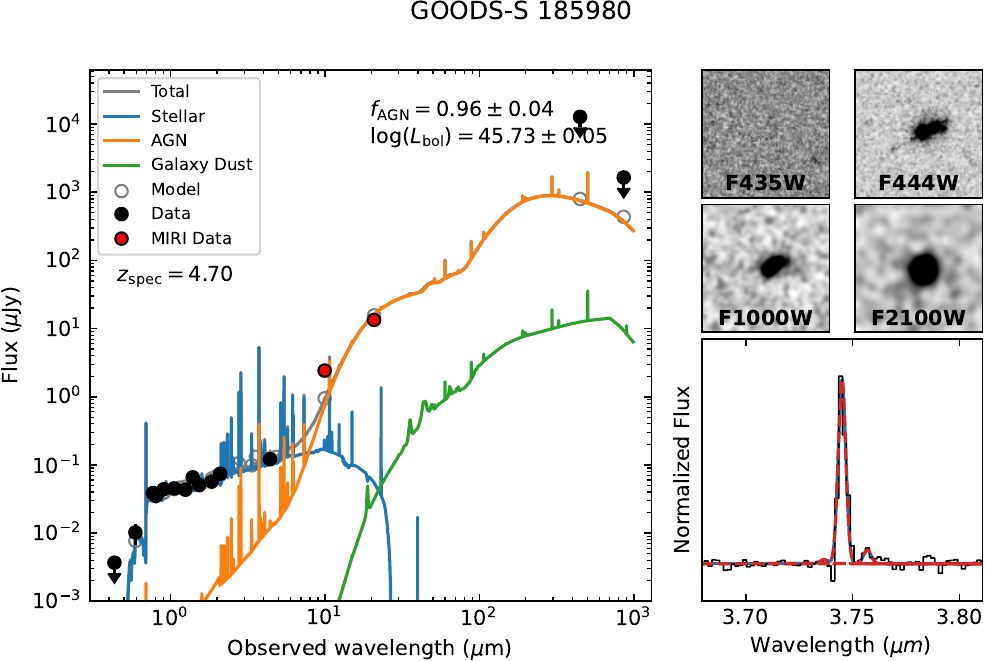}

		\caption{SED, image cutouts and spectra of the five narrow-line AGN in our sample. The left panel of each object shows the observed MIRI photometry as red points with black outlines, HST, JWST/NIRCam and SCUBA-2 photometry as black points, and the 2$\sigma$ upper limits for nondetections. The best-fit model is shown in gray, the galaxy stellar and nebular emission in blue, the AGN emission in orange, and the host galaxy dust in green. The top right panel shows the 5'' stamp images in the ACS F435W, NIRCam F444W and MIRI F1000W and F2100W filters. ACS F435 is a dropout band at our target redshift range. The bottom right panel shows the H$\alpha$ emission line spectrum from the NIRCam grism or NIRSpec. The black histograms show the observed spectra, the blue line the total best-fit model, and the red dashed lines the narrow components.}
		\label{fig:sed1}
\end{figure*}

\begin{figure*}[!t]
	\centering
        \includegraphics[width=0.49\textwidth]{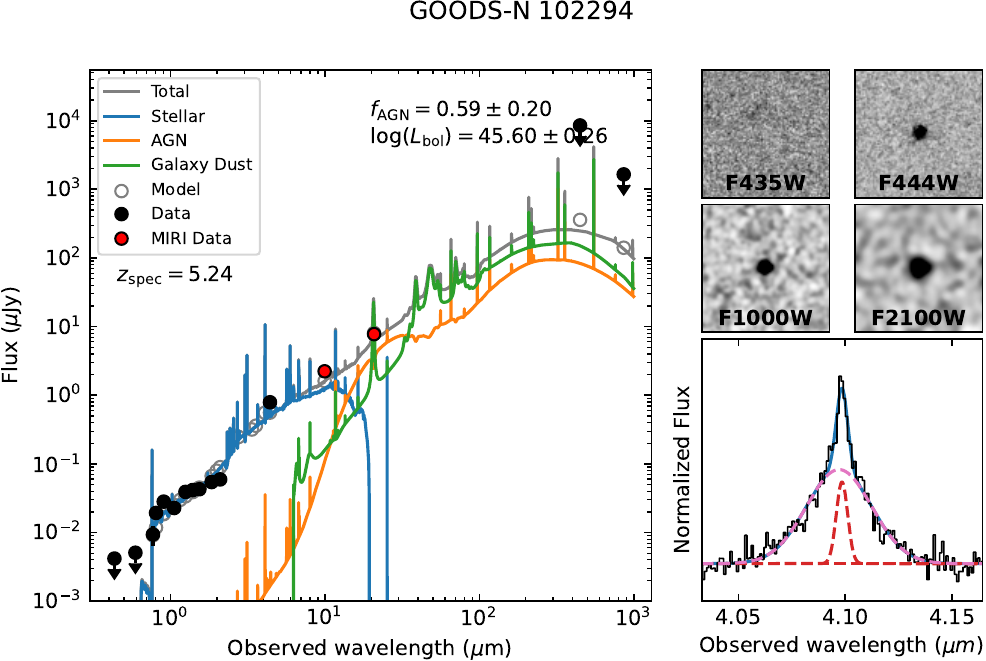}
        \hfill
        \includegraphics[width=0.49\textwidth]{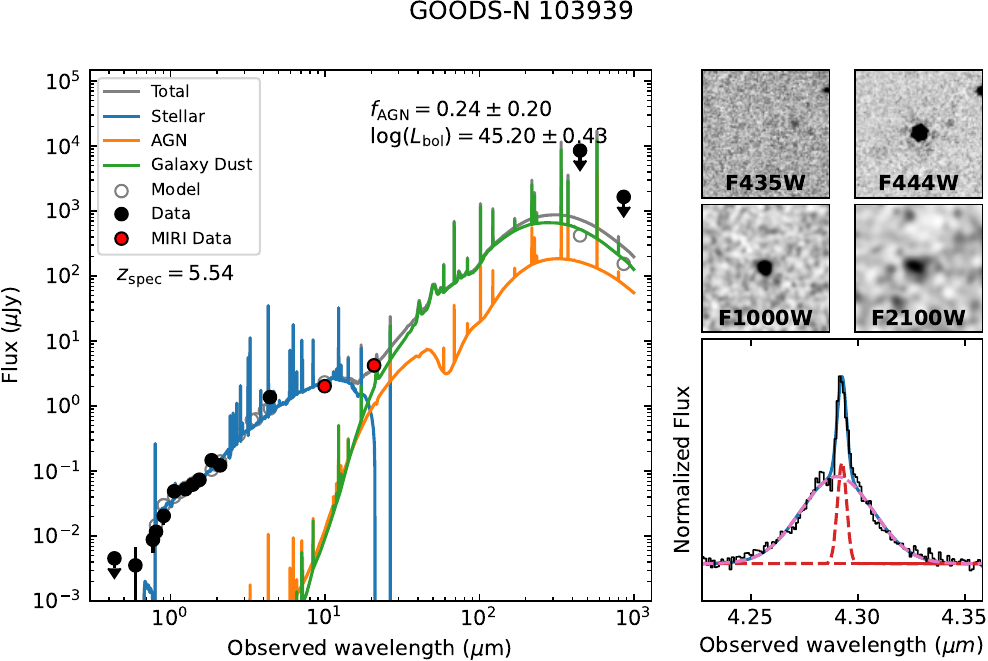}

		\caption{Same as Figure \ref{fig:sed1}, but for the two BL AGN. The broad component is shown in the magenta dashed line.}
		\label{fig:sed2}
\end{figure*}

We compare the observed colors to AGN and galaxy templates from the literature, redshifted to $z=4-6$. These include templates of featureless AGN, silicate AGN, and star-forming galaxies constructed from deep Spitzer IRS spectroscopy in the GOODS-Herschel \citep{Kirkpatrick12} and CEERS fields \citep{Kirkpatrick23}.
All objects in our sample with $\fagn > 0.2$ overlap with the AGN templates in color-color space, but not with star-forming galaxies. 
One object with $0.1 < \fagn < 0.2 $ occupies color-color space between that of AGN and SFGs. While its colors are close to those of SFGs, it is redder in F1000W-F2100W.
In the right panel of Figure \ref{fig:color}, we show the observed F210M, F444W, F1000W and F2100W photometry of our sample, color-coded by \fagn \ and overplotted on the AGN and galaxy templates. We observe a strong agreement between the AGN templates and the SED of objects with $\fagn \gtrsim 0.2$. 
Two objects with $\fagn \approx 0.1-0.2$ lie between the AGN and SFG templates.
Overall, \fagn \ provides a strong proxy for separating AGN and galaxy SEDs.

Following \citet{Yang23}, we adopt a three-tier classification scheme based on the \fagn \ value, categorizing objects into pure AGN, composite AGN-star-forming systems, and star-forming galaxies (hereafter AGN, composite and SF). We adopt the following classification:
(1) AGN: $\fagn \ge 0.5$,
(2) Composite: $0.1 \le \fagn < 0.5$, and
(3) SF: $\fagn < 0.1$.
This returns 10 AGN, six composite, and no SF systems at $z \gtrsim 5$. Our final sample spans $z=4.5-7.2$, and $\lbol=10^{44.6-46.8}\ \mathrm{erg~s}^{-1}$. We show the distribution of the sample in redshift and \fagn \ in Figure \ref{fig:sample}. We present tabulated data for the full AGN sample at $z \gtrsim 5$ in Table \ref{tab:sample}.

\begin{deluxetable}{lccccccc}
\tablecaption{AGN Sample at $z \gtrsim 5$\label{tab:sample}}
\tablehead{
\colhead{ID} & \colhead{R.A.} & \colhead{Decl.} & \colhead{$z$} & \colhead{Spec-$z$} & \colhead{$\log({\lbol}/\mathrm{erg~s}^{-1})$} & \colhead{$f_{\rm AGN}$} & \colhead{Alternative IDs} \\
\colhead{(1)} & \colhead{(2)} & \colhead{(3)} & \colhead{(4)} & \colhead{(5)} & \colhead{(6)} & \colhead{(7)} & \colhead{(8)}
}
\startdata
GN-8224 & 189.139221 & 62.235697 & 5.303 & Y & $46.23 \pm 0.14$ & $0.66 \pm 0.10$ & N-7496, S2, N2GN\_1\_01, GN10 \\
GN-11419 & 189.287911 & 62.217458 & 5.10 & N & $45.08 \pm 0.44$ & $0.29 \pm 0.23$ & \nodata \\
GN-13928 & 189.235613 & 62.202050 & 5.182 & Y & $46.43 \pm 0.14$ & $0.65 \pm 0.10$ & N-2663, S3, N2GN\_1\_23, GN15 \\
GN-96071 & 189.105789 & 62.188837 & 5.162 & Y & $45.36 \pm 0.29$ & $0.56 \pm 0.25$ & \nodata \\
GN-102294 & 189.344277 & 62.263369 & 5.24 & Y & $45.60 \pm 0.26$ & $0.59 \pm 0.20$ & GN-12839, N-12839, GN-1088832 \\
GN-102397 & 189.326121 & 62.263051 & 4.54 & N & $45.54 \pm 0.18$ & $0.56 \pm 0.15$ & \nodata \\
GN-103939 & 189.281012 & 62.247309 & 5.538 & Y & $45.20 \pm 0.43$ & $0.24 \pm 0.20$ & GN-9771, N-9771, GN-1087388 \\
GN-104054 & 189.019245 & 62.243530 & 7.041 & Y & $46.09 \pm 0.26$ & $0.60 \pm 0.15$ & N-9094 \\
GN-107919 & 189.070649 & 62.208948 & 7.1868 & Y & $45.97 \pm 0.02$ & $0.80 \pm 0.00$ & GNz7q \\
GN-104946 & 189.320410 & 62.233585 & 5.188 & Y & $45.19 \pm 0.42$ & $0.25 \pm 0.21$ & N-7162 \\
GN-903 & 189.198327 & 62.297038 & 7.0337 & Y & $45.88 \pm 0.33$ & $0.66 \pm 0.20$ & \nodata \\
GN-6404 & 189.244312 & 62.247537 & 5.215 & Y & $44.64 \pm 0.54$ & $0.12 \pm 0.12$ & N2GN\_1\_13, GN32 \\
GS-18381 & 53.069023 & -27.807192 & 4.99 & Y & $45.65 \pm 0.28$ & $0.44 \pm 0.19$ & \nodata \\
GS-20329 & 53.067411 & -27.812397 & 4.56 & N & $45.59 \pm 0.08$ & $0.97 \pm 0.04$ & \nodata \\
GS-39981 & 53.094685 & -27.865169 & 4.86 & N & $44.96 \pm 0.41$ & $0.31 \pm 0.22$ & \nodata \\
GS-185980 & 53.097177 & -27.768770 & 4.703 & Y & $45.73 \pm 0.05$ & $0.96 \pm 0.04$ & \nodata \\
\enddata
\tablecomments{Columns (2) and (3): coordinates in decimal degrees. Column (5): Y if redshift is spectroscopic; N if otherwise. Column (8): alternative IDs in the literature, taken from \citet{Pope05}, \citet{Fujimoto22}, \citet{Matthee24}, \citet{Xiao24}, \citet{Xiao25}, \citet{Lagache26}, and \citet{Zhang26}.}
\end{deluxetable}

\subsection{Narrow-line AGN}

NIRCam F444W grism spectra are available from the FRESCO survey for seven of the AGN at $z=5.1-5.6$, providing coverage of the H$\alpha$ emission line at these redshifts. This allows us to examine the presence or absence of broad emission lines in these objects. \citet{Matthee24} has performed a search for BL AGN using FRESCO spectra in GOODS-N and GOODS-S, while \citet{Zhang26} has performed an independent search in GOODS-N using FRESCO and JADES spectra. We cross-match our MIRI-selected AGN to the BL AGN in these two studies. 

We find that five out of the seven AGN within F444W grism coverage do not have broad H$\alpha$ emission lines reported in \citet{Matthee24}
and \citet{Zhang26}. These five AGN represent narrow-line, obscured AGN only detectable by their IR hot dust signatures observed by MIRI. Among these five objects, four of them have $\fagn > 0.5$, placing them in our pure AGN classification, while the remaining one is a composite object with $0.1 < \fagn < 0.5$. We show the SEDs, image cutouts, and spectra for these objects in Figure \ref{fig:sed1}

Two of these narrow-line AGN, GOODS-N 8224 and GOODS-N 13928, are known submillimeter galaxies in the field, first identified by \citet{Pope05} as GN10 and GN15, respectively. Both objects are additionally identified as potential ultra-massive galaxies in \citet{Xiao24}, where they were referred to as S2 and S3. Despite the lack of broad emission lines, our SED modeling suggests a high AGN contribution in these two objects, with $\fagn \sim 0.6-0.7$, and a bolometric luminosity of $\lbol \sim 10^{46} \ \mathrm{erg~s}^{-1}$.

\subsection{BL AGN}\label{sec:bl}

Apart from the narrow-line AGN, two of our MIRI-selected AGN, GOODS-N 102294 and GOODS-N 103939, have reported broad line detections in \citet{Matthee24} and \citet{Zhang26}. They correspond to GOODS-N-9771 and GOODS-N-12839 in \citet{Matthee24} and GN-1087388 and GN-1088832 in \citet{Zhang26}, respectively. We show their SEDs, image cutouts, and spectra in Figure \ref{fig:sed2}.

GOODS-N 103939 shows an absorption line in its broad H$\alpha$ emission line, a signature of LRDs. Our SED modeling returns a \fagn \ of 0.24, placing it in the composite category. On the other hand, GOODS-N 102294 shows no clear absorption features in its H$\alpha$ emission line \citep[see also][]{Matthee26}. Interestingly, our SED modeling measures a \fagn \ of 0.59, classifying it as a pure AGN.

\begin{figure*}[!t]
	\centering
        \includegraphics[width=0.32\textwidth]{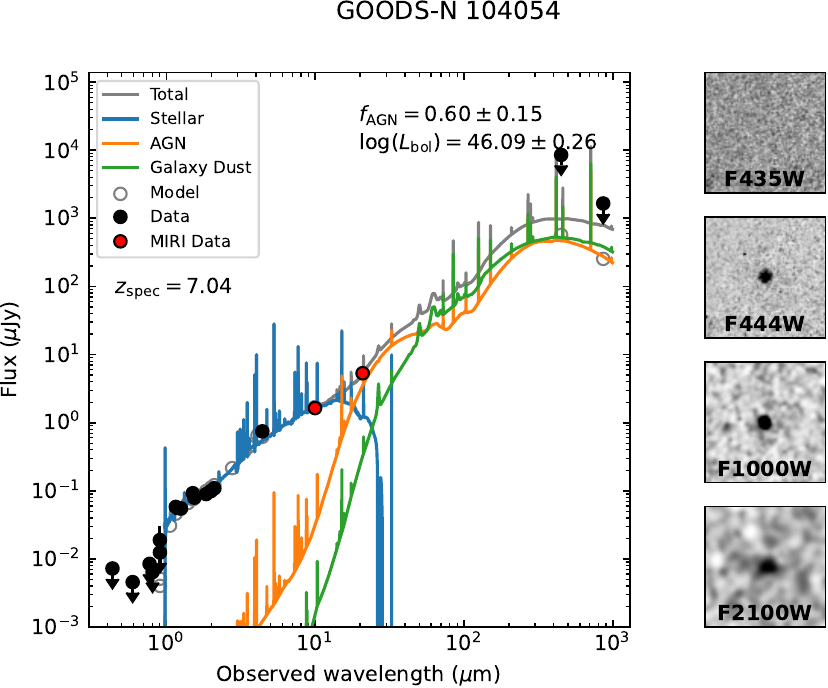}
        \includegraphics[width=0.32\textwidth]{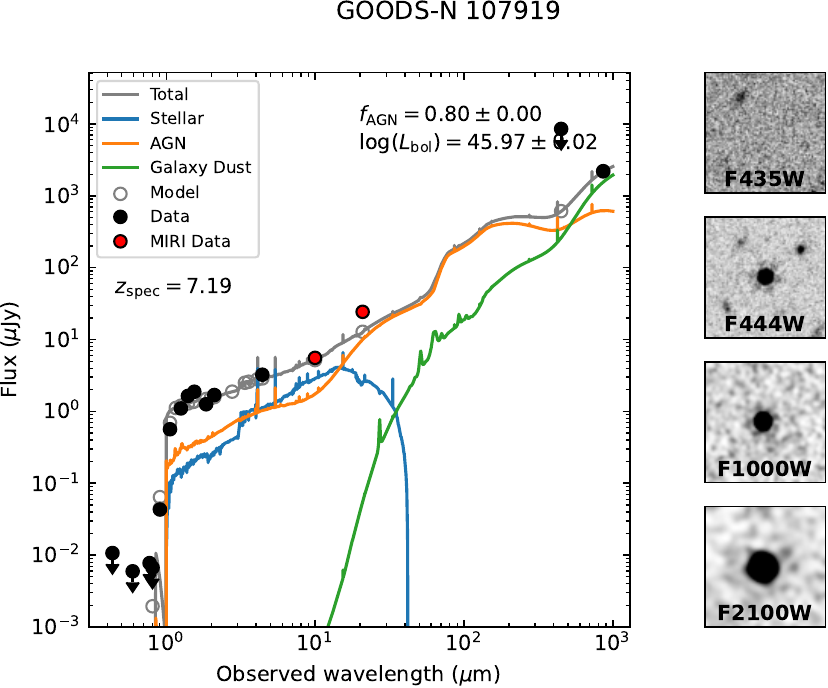}
        \includegraphics[width=0.32\textwidth]{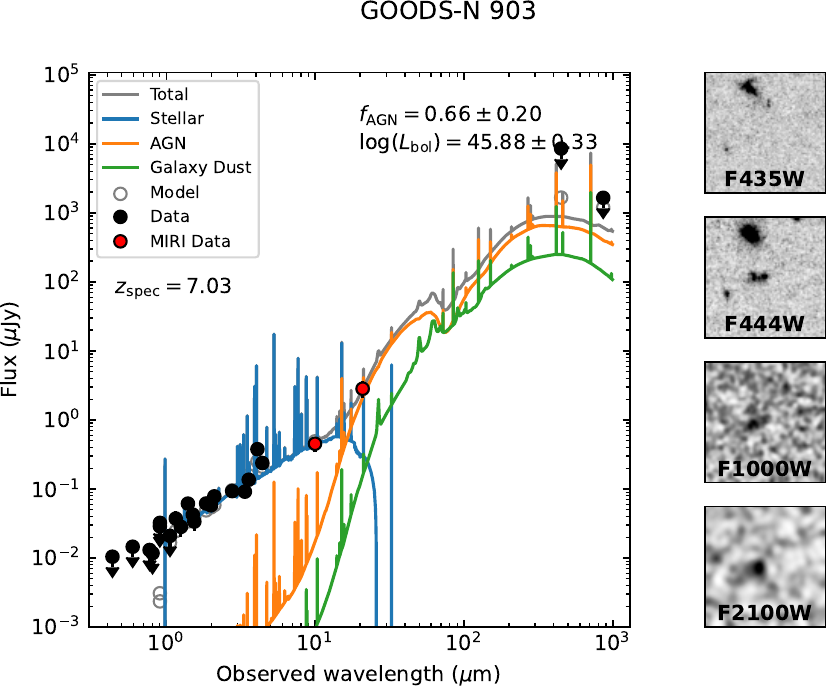}

		\caption{Same as Figure \ref{fig:sed1}, but for the $z>7$ AGN.}
		\label{fig:sed3}
\end{figure*}

Both of these sources fall outside of JADES F277W and F356W imaging coverage, preventing a conventional photometric LRD classification. The low \fagn \ of GOODS-N 103939, resulting from its flatter MIRI colors, is consistent with the weak MIRI emission observed in LRDs \citep[e.g.][]{Williams24, Perez-Gonzalez24, Leung25}. Nonetheless, the MIRI detection in this object, albeit weak, is consistent with evidence that some rest-frame infrared emission, potentially due to warm dust, is present in LRDs \citep[e.g.][]{Leung25, Delvecchio25, Barro26, Ronayne26}. In contrast, the high \fagn \ of GOODS-N 102294 suggests that it could be consistent with a typical AGN.  Alternatively, if GOODS-N 102294 is indeed an LRD, its rest-frame near-infrared color would be one of the reddest measured for the LRDs, potentially suggesting unusually strong hot-dust emission.

\subsection{$z>7$ AGN}
We also highlight the three AGN in our sample at $z > 7$, GOODS-N 903, 107919, and 104054. All three of these AGN have confirmed spectroscopic redshifts in the FRESCO [OIII] emitter catalog. Their SEDs and image cutouts are shown in Figure \ref{fig:sed3}.

The AGN GOODS-N 107919 corresponds to the $z=7.1899$ red quasar GNz7q, first reported in \cite{Fujimoto22}. This quasar is selected in our SED fitting with an \fagn \ of 0.8 and \lbol \ of $10^{46}\ \mathrm{erg~s}^{-1}$. A joint spectrophotometric analysis using MEOW data and other JWST NIRCam, MIRI and NIRSpec data is presented in \citet{Fei26}.

The AGN GOODS-N 104054 was presented as ID9094 in \citet{Xiao25} as a BL AGN through its H$\beta$ emission line and a potential LRD. That study reported NOEMA non-detections in [CII] and 1.3 mm, arguing for an AGN origin for this source. Using our MIRI data, we find a high \fagn \ of 0.6 for this source, showing strong hot dust emission due to the AGN. Similar to the $z\sim 5$ BL AGN GOODS-N 102294 in the previous section, this object likely represents a typical dust-obscured AGN.

In addition to these two known AGN, we identify GOODS-N 903 from our SED analysis. This object is at a spectroscopic redshift of $z=7.03$, and displays strong hot dust emission with \fagn=0.66. Interestingly, it displays a clumpy morphology in its NIRCam imaging, but appears compact in F2100W. This is consistent with the picture of a dust-obscured AGN residing in an extended galaxy, where the circumnuclear dust emission dominates the F2100W band and the host galaxy dominates the NIRCam bands.

\begin{deluxetable}{ccccc}
\tablecaption{Bolometric Luminosity Function at $z=4.5$--$6$}\label{tab:lf}
\tablehead{\colhead{$\log({\lbol}/\mathrm{erg~s}^{-1})$} & \colhead{$\phi \times 10^{-5}$} & \colhead{$N_\mathrm{AGN}$} & \colhead{$V_\mathrm{eff}$} \\
\colhead{} & \colhead{(Mpc$^{-3}$mag$^{-1}$)} & \colhead{} & \colhead{(Mpc$^{3}$)}}

\startdata
44.55 & $2.716_{-1.808}^{+2.688}$ & 1 & 105287 \\
45.25 & $1.978_{-0.819}^{+1.065}$ & 7 & 426765 \\
45.95 & $1.299_{-0.604}^{+0.815}$ & 4 & 518525 \\
46.65 & $0.476_{-0.300}^{+0.518}$ & 1 & 522886 \\
47.35 & $<0.30$ & 0 & 537773
\enddata
\end{deluxetable}\label{tab:lf}

\subsection{Bolometric Luminosity Function}

A key diagnostic to determine the importance of the MIRI-selected AGN in early SMBH growth is the bolometric luminosity function. We construct the AGN bolometric luminosity function in the redshift range of $z=4.5-6.0$. We adopt the AGN bolometric luminosity from the SED fitting results with \texttt{CIGALE} using the \texttt{accretion\_power} parameter, which represents the angle-averaged intrinsic AGN disk luminosity prior to dust attenuation. 

We first calculate the effective volume as a function of \lbol , for the AGN and composite sources, respectively. 
The effective volume is the integral of the completeness function over the comoving volume element,
\begin{equation}
V_\mathrm{eff}(\lbol) = \int \frac{dV}{dz} C(\lbol,z) dz,
\end{equation}
where $C(\lbol,z)$ is the completeness calculated in Section \ref{sec:samp_comp}. This is calculated independently for the AGN and composite sources. The resulting effective volumes are shown in the right panel of Figure \ref{fig:volume}. We then calculate the final effective volume for each luminosity bin by weighting the number of sources in the AGN and composite categories within that luminosity bin as
\begin{equation}
V_\mathrm{eff, final} = \frac{V_\mathrm{eff, AGN} N_\mathrm{AGN} + V_\mathrm{eff, mix} N_\mathrm{mix}}{N_\mathrm{AGN}+N_\mathrm{mix}}.
\end{equation}

We calculate the luminosity function following the methodology of \citet{Leung23}, using a Markov Chain Monte Carlo (MCMC) method. We adopt a uniform prior in number density and a Poisson likelihood function. Within each step of the MCMC chain, we randomly resample the luminosity of each object in a normal distribution centered at the best-fit \lbol \ with a standard deviation equal to its uncertainty, allowing an object to move between luminosity bins within each MCMC step. This procedure accounts for uncertainties from both Poisson statistics and the uncertainty in \lbol . We adopt the posterior median as the bolometric luminosity function value and the 16th and 84th percentiles as the uncertainty. For luminosity bins with no detected objects, we adopt the 95th percentile as the upper limit. We report the luminosity function in Table \ref{tab:lf}.

\begin{figure*}[!t]
	\centering
		\includegraphics[width=0.7\textwidth]{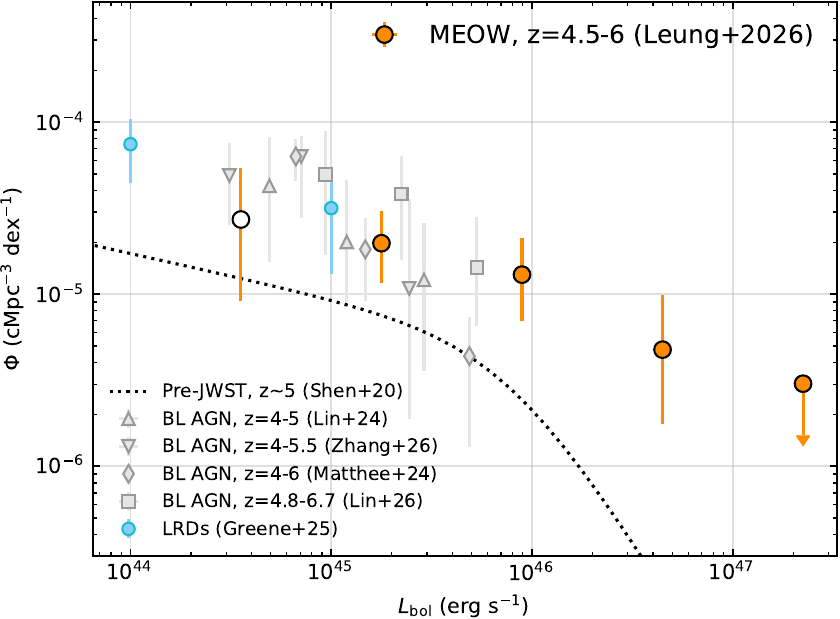}
		\caption{The bolometric luminosity function at $z=4.5-6$. The orange points show the number density calculated from our sample, with 68\% upper limits shown for luminosity bins with no detected objects. The unfilled circle denotes luminosity bins which are $<20\%$ complete. Luminosity functions of BL AGN from \citet{Lin24}, \citet{Matthee24}, \citet{Zhang26} and \citet{Lin26} are shown in gray symbols. The LRD luminosity function from \citet{Greene26} is shown as blue points. The pre-JWST luminosity function based on the compilation of \citet{Shen20} is shown by the black dashed line. Our MIRI-selected AGN show comparable number densities to both BL AGN and LRDs, and exceed pre-JWST results.}
		\label{fig:lf}
\end{figure*}

We plot our bolometric luminosity function at $z-4.5-6$ in Figure \ref{fig:lf}. We also show luminosity functions at $z\sim 5$ of BL AGN identified in JWST data from \citet{Matthee24}, \citet{Lin24,Lin26} and \citet{Zhang26}, as well as the pre-JWST AGN luminosity function in \citet{Shen20}. Compared to pre-JWST estimates, the number density of the MIRI-selected AGN is higher by $\sim 2$ times at $\lbol \sim 10^{45} \ \mathrm{erg~s}^{-1}$, rising to $\gtrsim 5$ times at $\lbol \sim 10^{45} \ \mathrm{erg~s}^{-1}$. The observed number density of our MIRI-selected AGN is broadly comparable to that of BL selected ones in different JWST datasets in the \lbol \ range of $\sim 10^{45} \ \mathrm{erg~s}^{-1}$, while the MIRI-selected AGN outnumber the BL ones at $\lbol \gtrsim 10^{46} \ \mathrm{erg~s}^{-1}$. For comparison, we also plot the luminosity function at $z=4-6$ of LRDs from \citet{Greene26} using empirically measured luminosities in Figure \ref{fig:lf}. The luminosities of LRDs cover the faint end of our MIRI-selected AGN, and the number densities are comparable between the two populations.

\section{Discussion}\label{sec:discussion}

\subsection{Comparison to other MIRI-selected AGN Samples}

Our analysis in the GOODS-S field encompasses areas covered by SMILES, which has conducted MIRI observations in eight MIRI filters in an adjacent, non-overlapping region with the MEOW footprint. Data from both surveys are included in our analysis to form a contiguous area within GOODS-S. \citet{Lyu24} selected 15 AGN candidates at $z>4.5$ in the SMILES footprint by SED modeling using HST, NIRCam and MIRI photometry. The main difference between the two studies is the MIRI filters included. While \citet{Lyu24} used all eight available MIRI filters in the SMILES data, including F1000W and F2100W, we only included the latter two filters for homogeneous filter coverage with the MEOW observations in the rest of the field. Among the additional filters in SMILES, they reach $5\sigma$ depths of $\sim 0.2$ $\mu$Jy in F550W and F770W, and $\sim 0.6$ to 1.8 $\mu$Jy in F1280W, F1500W and F1800W. By comparison, MEOW reaches $5\sigma$ depths of 0.5 and 3.6 $\mu$Jy in F1000W and F2100W, respectively.

None of the 15 sources are selected in our sample. Among the 15, eight are undetected in both the F1000W and F2100W filters based on our source detection, while six are detected in F1000W but not F2100W. Their inclusion in \citet{Lyu24} is likely based on detections in the additional filters in SMILES which are not used in our analysis. Eight of the undetected sources have an \lbol \ lower than $10^{45} \ \mathrm{erg~s}^{-1}$ in \citet{Lyu24}, where our completeness drops to $\lesssim 20\%$ (see Figure \ref{fig:volume}). One of the SMILES sources is detected in both F1000W and F2100W based on our source detection, but our photometric redshift estimation prefers a low-redshift solution of $z=2.4$.
Our sample includes one AGN candidate within the SMILES footprint that is not selected by \citet{Lyu24}. This is GOODS-S 39981, which is classified as a composite source with an \fagn \ of $0.3 \pm 0.2$ and \lbol \ of $10^{45\pm 0.4} \ \mathrm{erg~s}^{-1}$ in our analysis.

\begin{figure}[!t]
	\centering
		\includegraphics[width=0.45\textwidth]{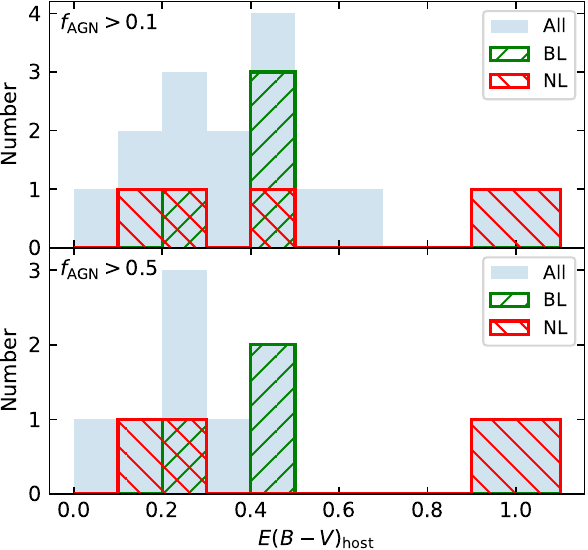}
		\caption{Distribution of the host $E(B-V)$ from our SED modeling. The top panel shows all the objects in our sample, while the bottom panel shows those with $\fagn > 0.5$. The blue filled histogram shows all the AGN, the green filled histogram the BL AGN, and the red filled histogram the narrow-line AGN. The narrow-line AGN span a wide range of host extinction from $\sim 0.1$ to $\sim 1$, indicating circumnuclear obscuration as well as potential obscuration by galaxy-scale dust.}
		\label{fig:ebv}
\end{figure}

\subsection{The Source of Obscuration}

With a substantial population of obscured AGN detected by MIRI with no broad emission line signatures at $z \gtrsim 5$, it is imperative to understand the source of obscuration in these systems. In the unified AGN model \citep{Antonucci93, Urry95}, the distinction between an obscured and unobscured AGN lies in the inclination angle of a circumnuclear dusty torus relative to the observer. However, more recent evidence suggests that the assumption of a single smooth torus is oversimplified, and the structure of the circumnuclear dust is likely represented by a more complex, clumpy distribution, consisting of both an equatorial and polar component \citep[e.g.][]{Netzer15,RamosAlmeida17}. Alternatively, obscuration due to materials in the host galaxy on larger scales has long been known to affect AGN in the local universe \citep[e.g.][]{Lawrence82, Malkan98, Goulding12, Wu22}. Analyses of the evolution of the interstellar medium (ISM) suggest that obscuration by galaxy-scale materials could play an increasingly prominent role among AGN at high redshift \citep[e.g.][]{Gilli22}. 

We investigate this scenario by examining the degree of dust obscuration in the host galaxy for our AGN sample, measured by the host $E(B-V)$ in our SED model. While this parameter measures the obscuration of the galaxy's stellar emission rather than the AGN, it is a useful proxy for the overall dust content in the host galaxy and thus the environment in which these AGN reside. We show the distribution of the host $E(B-V)$ in Figure \ref{fig:ebv} for the full AGN sample, and the sources with confirmed broad or narrow emission lines. The full AGN sample spans a wide range of host galaxy obscuration, with $E(B-V)\sim 0.1 - 1$, indicating that the MIRI-selected AGN reside in diverse environments in terms of host dust content. 

Of particular interest are the five AGN with narrow emission lines, which also span a wide range of $E(B-V)$ from 0.2 to 1. For those with the lowest $E(B-V)$, the modest dust content in the host implies that the obscuration likely comes from the circumnuclear torus. Two of the narrow-line AGN display the highest $E(B-V)$ of $\approx 1$ among the full AGN sample, indicating that these AGN live in a dusty host environment. Host galaxy obscuration could plausibly play a role in these systems, leading to some suppression of the rest-frame UV and optical continuum emission, as well as the broad optical line emission from the AGN. In the rest-frame near-IR, the hot circumnuclear dust emission can penetrate through the obscuring materials due to its longer wavelength, enabling detection by MIRI. Such differential galactic-scale obscuration has been observed in AGN at $z\sim 1-2$ \citep[e.g.][]{Silverman23}. Nonetheless, our current analysis cannot fully disentangle circumnuclear and galactic-scale obscuration in these sources. Finally, the BL AGN generally reside in host galaxies with moderate obscuration of $E(B-V) \approx 0.2-0.5$, suggesting lower but non-trivial obscuration in these systems.

\subsection{The Full Early AGN Population in Context}

\begin{figure}[!t]
	\centering
		\includegraphics[width=0.4\textwidth]{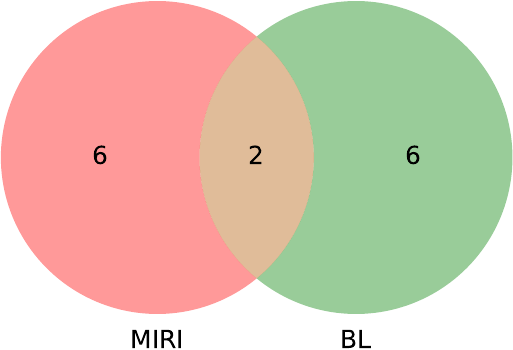}
		\caption{Venn diagram comparing the MIRI-selected AGN from this work and the BL AGN from \citet{Matthee24} within our footprint. The MIRI-selected AGN are restricted to those with spectroscopic redshifts of $4.9 \le z \le 6.6$, where the H$\alpha$ falls within the FRESCO grism spectra bandpass used by \citet{Matthee24}. The two samples are largely distinct, with only two objects simultaneously selected by MIRI and BL.}
		\label{fig:venn}
\end{figure}

In this section, we discuss our findings in the context of the complete SMBH population at high redshift. Our results have revealed a substantial population of obscured AGN at $z \gtrsim 5$ selected through their distinctive hot dust signature in the rest-frame near-IR. The luminosity function of these MIRI-selected AGN shows that they have comparable number densities as those selected by broad lines, down to $\log \lbol \sim 45$. Importantly, out of the seven MIRI-selected AGN where H$\alpha$ emission line spectra are available, only two show detectable broad emission lines, suggesting these are largely distinct populations (Figure \ref{fig:venn}). This highlights the necessity of MIRI observations in mapping the full AGN population at these redshifts. While current spectroscopic samples reach luminosities of $\log \lbol \sim 44.5$, the abundance of MIRI-selected AGN remains less certain at these lower luminosities given the current depth of our MIRI observations. One caveat of the BL luminosity functions at these redshifts is the presence of LRDs in BL samples. As the bolometric luminosities of BL AGN are calculated using canonical bolometric corrections, the luminosities of the LRDs in the sample are likely overestimated \citep{Greene26}, leading to an overestimation of the BL luminosity function. However, as MIRI selection is less sensitive to LRDs, the MIRI-derived luminosity function would better isolate the number density of typical AGN. As a result, the comparable number density of MIRI-selected AGN and BL AGN at face value, combined with the large fraction of narrow-line sources in our sample, implies that a substantial fraction of early SMBH growth could occur in dust enshrouded environments. A comprehensive analysis of the growing contribution of obscured AGN to cosmic SMBH growth toward higher redshifts is presented in \citet{Bulichi26}.

MIRI observations also serve as a useful tool to distinguish LRDs from typical AGN within BL selected objects. As shown in Section \ref{sec:bl}, BL objects display strong differences in emission in the MIRI wavelengths. Recent analysis suggest LRDs occupy the low luminosity regime of $\log \lbol \sim 43-45$ \citep{Greene26}. Interestingly, our luminosity function shows that the number density of MIRI-selected AGN is similar to that of high-luminosity LRDs at $\log \lbol \sim 45$. This could suggest that the typical hot-dust enshrouded accretion and the LRD mode are two pathways of early SMBH growth with comparable prevalence at these luminosities. At the lower luminosity regime where the bulk of the LRD population lie, we currently lack the sensitivity to make the same comparison. Extending the comparison between MIRI-bright obscured AGN and LRDs toward lower luminosities could provide insight into any evolutionary relation between the two populations.

In the future, deep MIRI observations will play a crucial role in completing the AGN census at $z \gtrsim 5$ by probing the lower luminosity regime. While NIRSpec observations can achieve high sensitivity for low-luminosity AGN, the need for target selection introduces selection biases. This selection effect can be overcome with NIRCam WFSS, but very deep grism observations will suffer from source confusion. By contrast, MIRI observations are able to capture both BL and narrow-line AGN without the pre-selection of targets, but could miss the LRDs which are weak in MIRI. Therefore, a multi-faceted approach is required to achieve a complete census of SMBHs in the early universe.

\section{Conclusions}\label{sec:conclusions}

We introduce the MEOW Survey, a MIRI imaging program in the GOODS-N and GOODS-S fields designed to detect dust obscured AGN through cosmic time, with a focus on $z \gtrsim 5$. MEOW is composed of 43 pointings covering 95 arcmin$^2$ of the GOODS-N and GOODS-S fields using the F1000W and F2100W filters, reaching $5\sigma$ photometric depths of 0.5 $\mu$Jy and 3.6 $\mu$Jy in the two filters. We present the sample of AGN at $z \gtrsim 5$ identified by SED modeling combining MEOW MIRI photometry with HST, NIRCam and SCUBA-2 data. Our main findings are summarized as follows.

\begin{enumerate}
    \item We present a sample of 16 MIRI-selected AGN at $z = 4.5$--7.2, of which 12 have spectroscopic redshift confirmations. The AGN span bolometric luminosities of $L_{\rm bol} = 10^{44.6}$--$10^{46.4}$~erg~s$^{-1}$.

    \item Of the 16 AGN, 12 are newly identified in this work, demonstrating the power of wide-area MIRI imaging to uncover AGN populations missed by previous surveys. Among these, 5 are narrow-line (Type II) AGN, representing the obscured population to which broad-line spectroscopic searches are insensitive.

    \item Two broad-line (Type I) AGN in our MIRI-selected sample exhibit markedly different mid-infrared emission properties. The object with modest MIRI emission is consistent with a little red dot (LRD), while the object with strong MIRI emission is consistent with either a typical AGN or an LRD with unusually strong hot-dust emission.

    \item We present an AGN bolometric luminosity function using our MIRI-selected sample at $z=4.5-6$. The resulting number densities exceed pre-JWST estimates, and are comparable to those of broad-line AGN and LRDs at similar redshifts, suggesting that the obscured AGN population contributes significantly to the total AGN census at $z \gtrsim 5$.

    \item The narrow-line AGN in our sample reside in diverse host environments, with evidence that some systems are subject to circumnuclear obscuration while others may be obscured on host-galaxy scales, pointing to multiple physical mechanisms driving AGN obscuration in the early universe.
\end{enumerate}

Our results demonstrate the unique power of JWST/MIRI in detecting and characterizing dust-obscured AGN in the early universe. This highlights that constructing a complete picture of early SMBH growth will require a multi-faceted approach combining MIRI imaging surveys and deep spectroscopic campaigns.

\begin{acknowledgments}
G.C.K.L., S.L.F., T.-E.B., and A.J.T. acknowledge support from NASA through STScI award JWST-GO-5407. Support for program \#5407 was provided by NASA through a grant from the Space Telescope Science Institute, which is operated by the Association of Universities for Research in Astronomy, Inc., under NASA contract NAS 5-03127. This work is based on observations made with the NASA/ESA/CSA James Webb Space Telescope. The data were obtained from the Mikulski Archive for Space Telescopes at the Space Telescope Science Institute, which is operated by the Association of Universities for Research in Astronomy, Inc., under NASA contract NAS 5-03127 for JWST.
\end{acknowledgments}

\vspace{5mm}
\facilities{JWST(MIRI, NIRCam, NIRSpec), HST(ACS, WFC3), SCUBA-2}

\software{\texttt{astropy} \citep{2013A&A...558A..33A,2018AJ....156..123A},
          \texttt{CIGALE} \citep{Boquien19, Yang20, Yang22}, \texttt{EAZY} \citep{Brammer08}, \texttt{emcee} \citep{Foreman-Mackey13},  
          \texttt{SExtractor} \citep{Bertin96}
          }

\appendix

\section{SED Fits for the Full Sample}

\begin{figure*}[ht]
	\centering
        \includegraphics[width=0.32\textwidth]{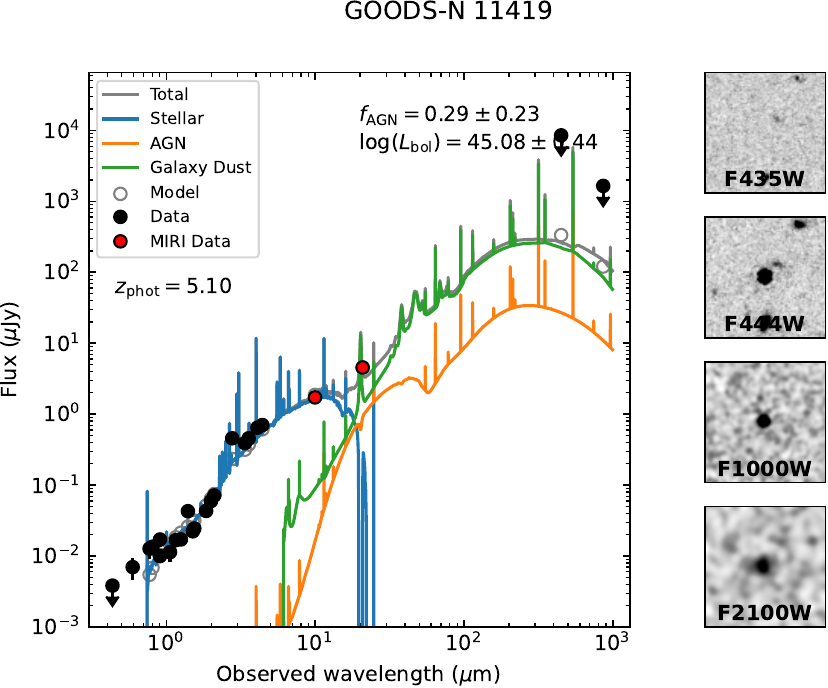}
        \includegraphics[width=0.32\textwidth]{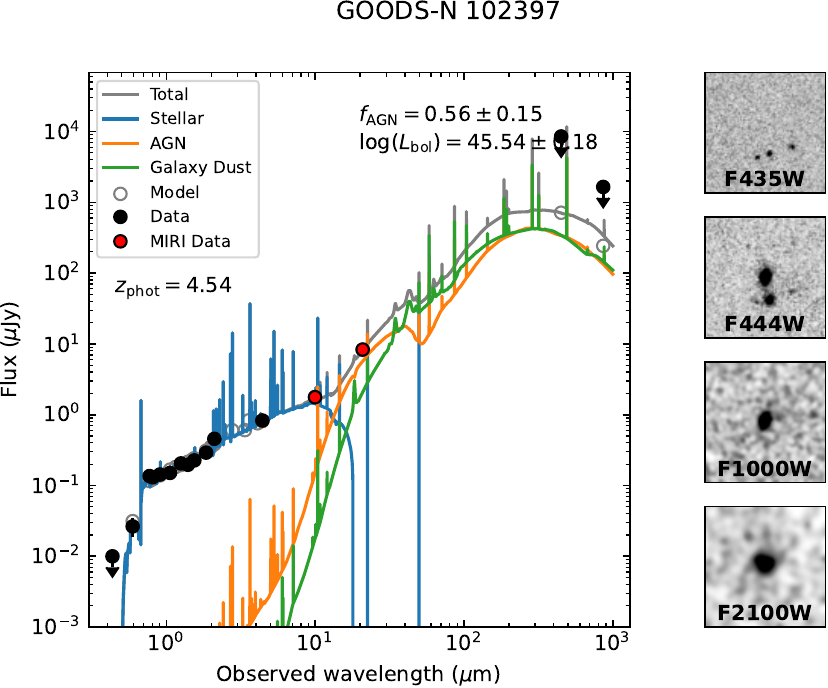}
        \includegraphics[width=0.32\textwidth]{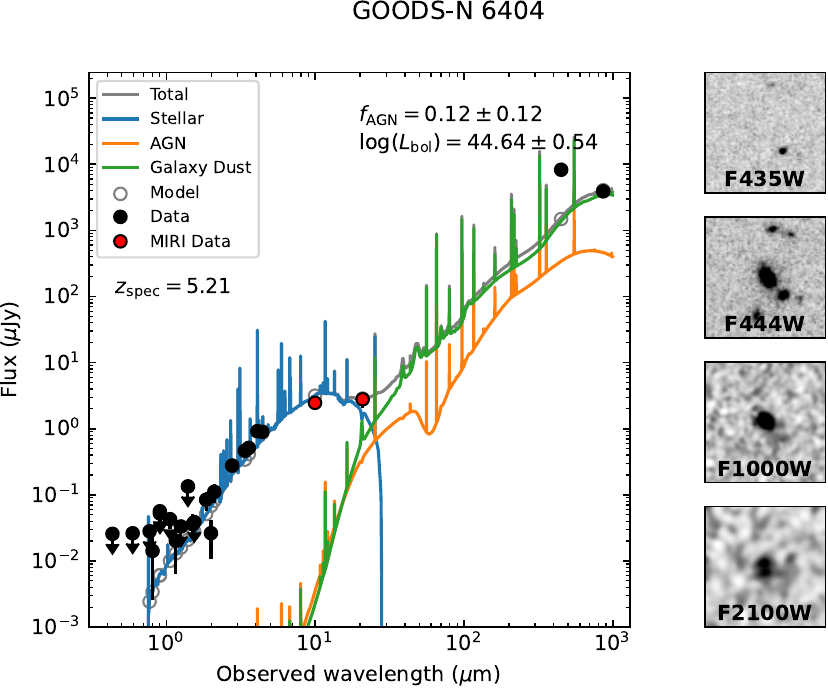}
        \\
        \includegraphics[width=0.32\textwidth]{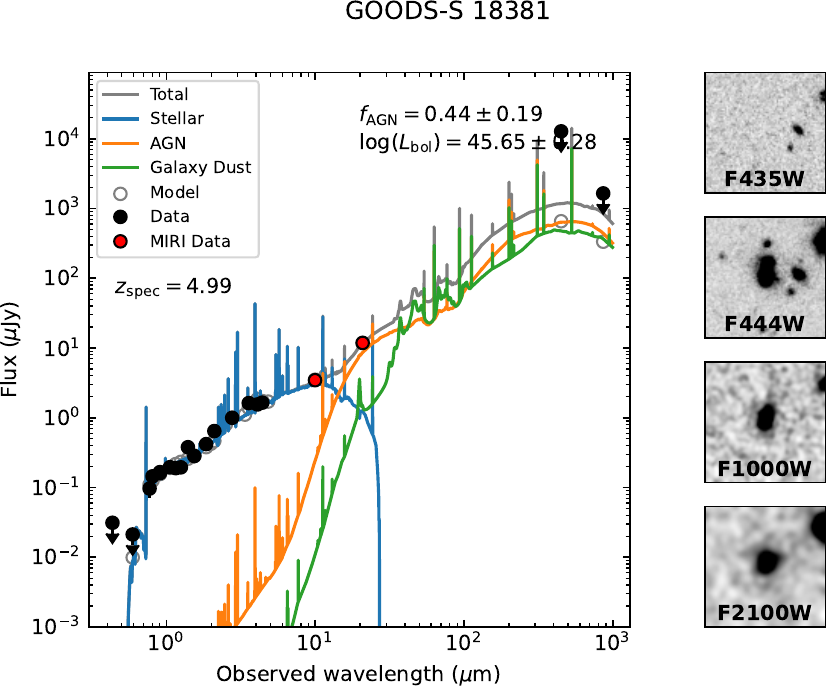}
        \includegraphics[width=0.32\textwidth]{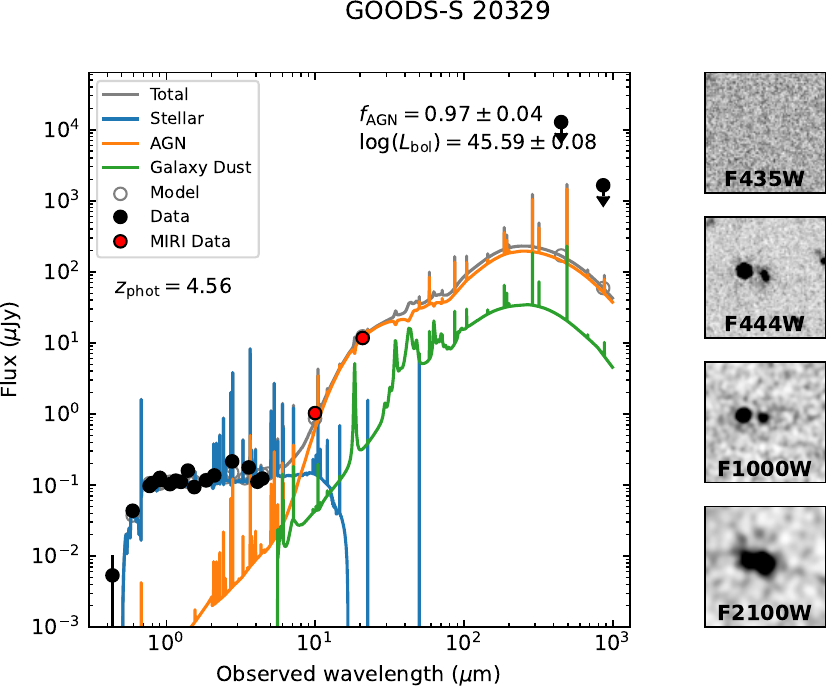}
        \includegraphics[width=0.32\textwidth]{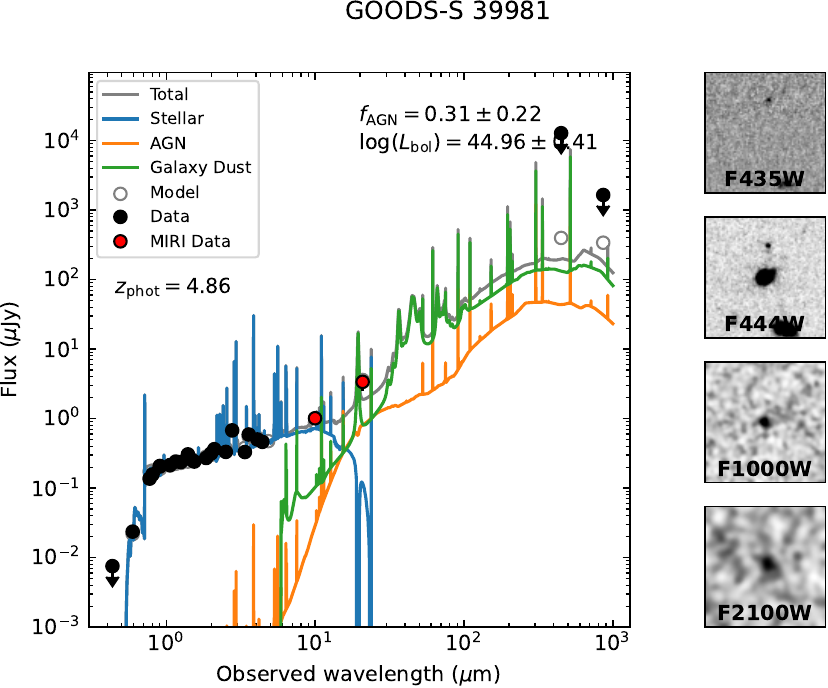}

		\caption{Same as Figure \ref{fig:sed1}, but for the remaining objects in the full sample.}
		\label{fig:sed4}
\end{figure*}

In Figure \ref{fig:sed4}, we show the SED and image cutouts for the remaining six AGN in our sample that have not appeared in the previous sections.

\section{Field-to-Field Variations in the Luminosity Function}

\begin{figure*}[!t]
	\centering
		\includegraphics[width=0.9\textwidth]{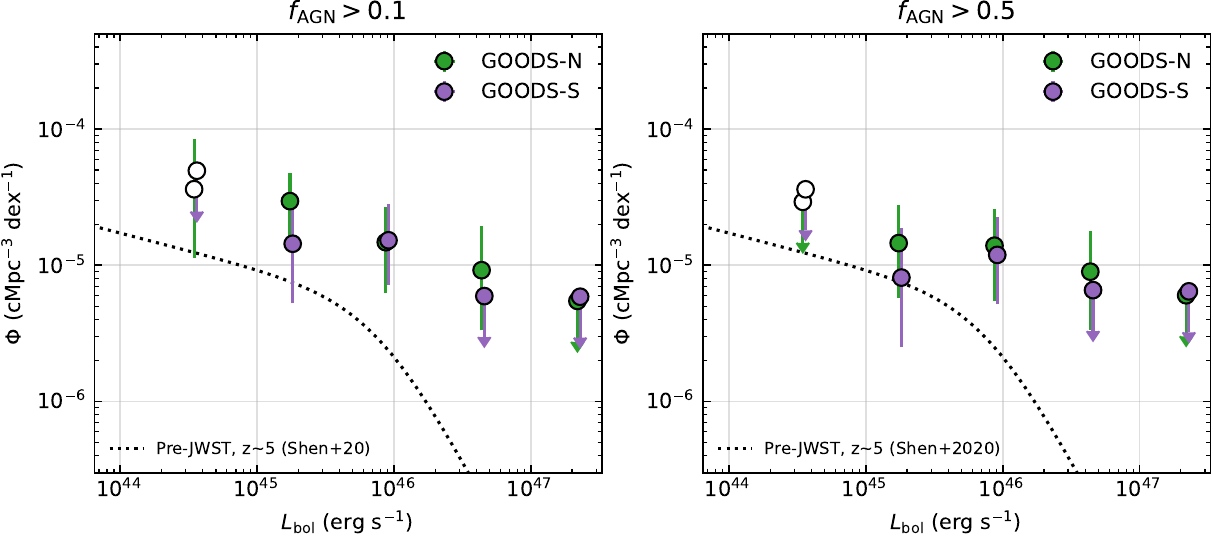}
		\caption{The luminosity function at $z=4.5-6$ for objects with $\fagn >0.1$ (left) and $\fagn > 0.5$ (right). Number densities in the GOODS-N and GOODS-S fields are shown by the green and purple points, respectively. Field-to-field variation is generally small, with GOODS-N showing a slightly higher density at $\lbol \gtrsim 10^{46} \ \mathrm{erg~s}^{-1}$. The inclusion of objects at $0.1 < \fagn < 0.5$ slightly increases the number density $\lbol \sim 10^{45} \ \mathrm{erg~s}^{-1}$ by a factor of $\sim 2$.}
		\label{fig:lf_fields}
\end{figure*}

In Figure \ref{fig:lf_fields}, we show the AGN bolometric luminosity function for sources with $\fagn  > 0.1$ and $\fagn  > 0.5$, with data in GOODS-N and GOODS-S plotted in different symbols. While we have more objects in GOODS-N than in GOODS-S, the field-to-field variation in the number density is within statistical uncertainty. The biggest difference appears to affect the higher luminosity bin at $\sim 10^{46.7} \ \mathrm{erg~s}^{-1}$, potentially suggesting that more luminous AGN are present in a known overdensity at $z\approx 5.2$ in GOODS-N \citep{Herard-Demanche25}. The number density in the moderate luminosity range ($\sim 10^{45-46} \ \mathrm{erg~s}^{-1}$) appears to be more consistent between the two fields. Finally, the inclusion of sources with \fagn \ $=$ 0.1--0.5 only affects the two lowest luminosity bins.

\bibliography{bibliography}{}
\bibliographystyle{aasjournalv7}



\end{document}